# Cathodoluminescence Analysis of Defects and Grain Boundaries in $Zn_3P_2$ Thin Films Grown on Graphene by MOVPE and MBE

Thomas Hagger[1], Mohammadreza Hassanzadeh[1], Aidas Urbonavicius[2], Ahmed El Alouani[3], Victor Boureau[4], Gulnaz Ganeeva[4], Nico Kawashima[5,6], Raphael Lemerle[1], Kamil Artur Wodzislawski[1], Sebastian Lehmann[2], Kimberly A. Dick[2], Silvana Botti[5,6], Adrien Michon[3], Anna Fontcuberta i Morral[1], and Simon Escobar Steinvall[2]

[1]Laboratory of Semiconductor Materials, Institute of Materials, School of Engineering, École Polytechnique Fédérale de Lausanne (EPFL), 1015 Lausanne, Switzerland

[2]Centre for Analysis and Synthesis and NanoLund, Lund University, Box 124, 221 00 Lund, Sweden

[3]Université Côte d'Azur, CNRS, CRHEA, Rue Bernard Grégory, 06560 Valbonne, France

[4]Interdisciplinary Center for Electron Microscopy (CIME), EPFL, Lausanne, Switzerland

[5]Research Center Future Energy Materials and Systems of the University Alliance Ruhr and ICAMS, Ruhr University Bochum, Universitätsstraße 150, D-44801 Bochum, Germany

[6]Institut für Festkörpertheorie und-optik, Friedrich-Schiller-Universität Jena, Max-Wien-Platz 1, 07743 Jena, Germany

Corresponding authors: anna.fontcuberta-morral@epfl.ch; simon.escobar_steinvall@chem.lu.se

**Abstract**

$Zn_3P_2$ is a promising earth-abundant absorber for thin-film photovoltaics, yet its development is hindered by the lack of lattice-matched substrates, its incompatible thermal expansion coefficient, and a complex defect landscape. Here, we demonstrate the quasi-van der Waals epitaxy of $Zn_3P_2$ on graphene by metal–organic vapour phase epitaxy (MOVPE) and directly link the density of antiphase boundaries to optical emission modulation using correlative electron microscopy and cathodoluminescence (CL). Moreover, it is observed through CL that grain boundaries act as non-radiative sinks for excited charge carriers. The effect extends several micrometres into the grains, making grain boundaries detrimental to the applicability of $Zn_3P_2$ in real devices. Further comparison with molecular beam epitaxy grown films reveals the suppression of strain-related sub-bandgap emission in MOVPE-grown material. Overall, quasi-van der Waals' epitaxy of $Zn_3P_2$ by MOVPE resulted in larger grains and improved material quality. In addition, these results directly link extended defects to recombination pathways in $Zn_3P_2$ and highlight grain-size control as a key strategy for improving earth-abundant photovoltaic absorbers.

## 1 Introduction

Established thin-film photovoltaic technologies such as CdTe and CIGS rely on elements that are either scarce or toxic, which may limit their long-term scalability.[1–6] Emerging alternatives such as perovskites and kesterites avoid some of these concerns but face other challenges, including stability, toxicity issues or complex defect chemistry.[7–9] This motivates the exploration of additional earth-abundant absorber materials for sustainable thin-film photovoltaic technologies.

One promising candidate is $Zn_3P_2$, a compound composed of earth-abundant elements and exhibiting favourable optoelectronic properties for photovoltaic applications. It possesses a direct bandgap of approximately 1.5 eV, close to the optimum for single-junction solar cells, and a high absorption coefficient exceeding $10^4$ $cm^{-1}$ in the visible spectral range.[10–13] In addition, long minority carrier diffusion lengths in the range of 5-7 $\mu$m have been reported.[14, 15] Early work on this material demonstrated a record conversion efficiency of nearly 6% in 1981.[16] Despite a theoretical detailed balance conversion efficiency limit close to 32%,[17] this efficiency has not been improved since, suggesting that fundamental material challenges remain.

One major challenge arises from the large lattice constants and thermal expansion coefficient of $Zn_3P_2$ relative to commonly used substrates, which complicates the growth of high-quality monocrystalline thin films.[18–20] High crystalline quality has been achieved using InP substrates, where monocrystalline films up to approximately 1 μm thickness have been reported,[21] and more recently, further improvements have been achieved using selective area epitaxy, where growth is initiated in nanoscale mask openings that allow strain relaxation before lateral overgrowth forms a continuous film.[22] However, these substrates currently rely on indium, a scarce element, which conflicts with the goal of developing a fully earth-abundant thin-film photovoltaic absorber.

Another challenge arises from the complex defect landscape of $Zn_3P_2$. The material is typically intrinsically p-type, and achieving stable n-type doping has proven difficult.[11, 23] First-principles calculations attribute this behaviour to the low formation energies of several intrinsic defects, which promote strong self-compensation and complicate compositional control.[24, 25] In addition to point defects, structural disorder such as strain fields and rotational domains has been reported in $Zn_3P_2$ thin films and can lead to spatial variations in optical emission.[26, 27] In many polycrystalline photovoltaic absorbers, such as Si, CdTe, CIGS, kesterites and hybrid perovskites, grain boundaries can strongly influence carrier recombination and transport.[28–36] However, the influence of grain boundaries on the optoelectronic properties of $Zn_3P_2$ remains poorly understood. Early work suggested that grain boundaries may not strongly limit device performance,[16] whereas later studies indicated that they can deteriorate electronic and optoelectronic properties.[37] Clarifying the role of extended defects such as grain boundaries is therefore important for understanding recombination processes and guiding the development of improved growth strategies.

Quasi–van der Waals epitaxy has emerged as a promising strategy by growing three-dimensional materials on two-dimensional substrates such as graphene.[38, 39] In this growth mode, the absence of strong covalent bonding at the interface relaxes lattice-matching constraints and can reduce strain during epitaxial growth. In addition, the weak interfacial bonding enables relatively straightforward exfoliation of the grown film and reuse of the substrate, which can reduce material consumption and improve the sustainability of the growth process.[40] Using this approach, Paul et al. demonstrated the growth of $Zn_3P_2$ on graphene by molecular beam epitaxy (MBE).[41] The resulting films exhibit a well-defined out-of-plane orientation but retain rotational freedom in the plane, leading to inherently polycrystalline thin films. Subsequent work showed that the grain size was limited to approximately 2 μm and strongly correlated with the quality of the graphene substrate, whereas variations in MBE growth parameters did not significantly increase the grain size.[27]

These observations suggest that the growth kinetics of MBE may inherently favour high nucleation densities, limiting the achievable grain size. This raises the question of whether alternative growth techniques could enable improved control over nucleation and grain size. A recent review on transition-metal dichalcogenide (TMD) growth highlighted that the grain size of two-dimensional materials can depend strongly on the growth technique.[42] In particular, MBE, which is a ballistic growth method, often results in higher nucleation densities and therefore smaller grain sizes compared to chemical vapour deposition (CVD) or metal–organic vapour phase epitaxy (MOVPE). These latter techniques typically operate at higher growth temperatures and involve surface-mediated precursor decomposition, which can increase surface diffusion lengths and favour the lateral growth of larger grains. This suggests that MOVPE may provide a promising route to achieve larger $Zn_3P_2$ grains on graphene.

In this work, we demonstrate the growth of $Zn_3P_2$ on graphene by MOVPE and investigate the relationship between structural defects and optical emission using combined electron microscopy and hyperspectral cathodoluminescence. By correlating antiphase boundary (APB) density and grain boundaries with spatially resolved luminescence, we provide direct insight into the recombination pathways governing polycrystalline $Zn_3P_2$ and evaluate the potential of MOVPE-grown material for photovoltaic applications.

## 2 Results and Discussion

### 2.1 MOVPE growth

$Zn_3P_2$ was successfully grown on graphene by MOVPE. In particular, the MOVPE growth results in substantially larger grains than previously reported for MBE-grown $Zn_3P_2$ on graphene.[27, 41] Representative scanning electron microscopy (SEM) images of two magnifications obtained under different pre-growth annealing and growth conditions are shown in Fig. 1 to showcase changes in grain size, morphology and nucleation density. The images were acquired near the centre of the substrate to minimise edge-related variations in growth conditions. While some spatial inhomogeneity was observed across the 1×1 $cm^2$ chips, the selected regions are representative of the dominant growth morphology.

The films consist of triangular grains, consistent with the intrinsic growth habit of $Zn_3P_2$ previously reported for growth on graphene.[27, 41] The grains exhibit a preferential top facet of a {101} plane while maintaining rotational freedom in-plane, resulting in polycrystalline films. A pronounced tendency for nucleation along linear features is visible in several samples. These lines correspond to the terraces of the underlying SiC substrate beneath the graphene, indicating that the substrate morphology influences the nucleation of $Zn_3P_2$ islands. Similar morphology-driven nucleation behaviour has also been observed in other van der Waals epitaxy systems.[27]

A strong influence of the pre-growth surface preparation on the resulting morphology is observed. Substrates treated with thermal annealing + phosphine ($PH_3$) prior to the growth show a higher density of misoriented or parasitic grains compared to the ones with an additional gaseous hydrochloric acid ($HCl_{(g)}$) flow during the thermal annealing. Increasing the growth temperature improves morphology, resulting in a greater number of better-defined triangular grains, especially when $HCl_{(g)}$ cleaning is included. Raman spectroscopy performed before and after growth (see Fig. S1 in the Supporting Information) confirms that the graphene layer remains intact, indicating that the $HCl_{(g)}$ treatment cleans the graphene surface without introducing detectable damage.

This behaviour indicates that surface contamination plays a key role in the nucleation of $Zn_3P_2$ on graphene. The observed reduction in nucleation density with *in-situ* $HCl_{(g)}$ treatment is therefore consistent with the removal of surface contaminants, resulting in a cleaner and more homogeneous graphene surface. As a consequence, a lower nucleation density enables the development of larger crystal grains. Further improvements in morphology are achieved by increasing the growth temperature in combination with the $HCl_{(g)}$ treatment, yielding substantially larger and more well-defined triangular grains. This behaviour can be attributed to enhanced adatom surface diffusion at elevated

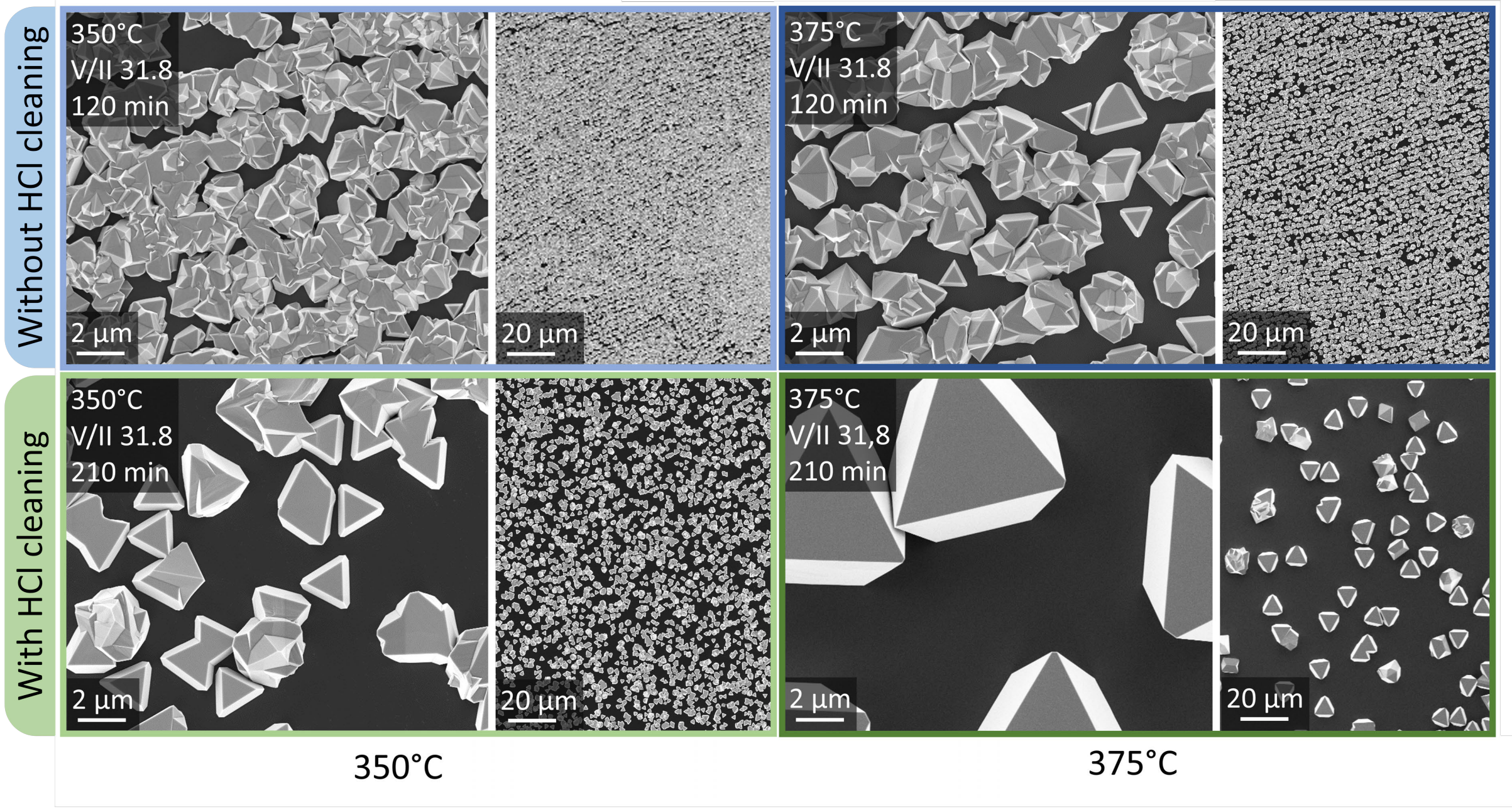

Figure 1: Two magnifications of top-view SEM images of $Zn_3P_2$ grown on graphene by MOVPE under different pre-growth annealing conditions and growth conditions. The top row corresponds to thermal annealing + $PH_3$ of the graphene substrate before the growth of $Zn_3P_2$, and the bottom row to thermal annealing + $PH_3$ + $HCl_{(g)}$. Growth conditions and times are indicated in the top-left corner. Scale bars: $2\mu$m and $20\mu$m

temperatures, and changes in the decomposition kinetics and surface transport of the metal–organic precursors and their intermediate species, which together influence the effective supply and incorporation of reactive species. Similar trends are observed across different graphene substrate qualities (see Fig. S2-S4), indicating that this behaviour is robust.

Because $Zn_3P_2$ growth on graphene results in polycrystalline films due to the rotational freedom of the grains, increasing the grain size is particularly important for potential device applications, as it reduces the density of grain boundaries that may negatively affect electronic and optoelectronic properties. In polycrystalline semiconductors, grain boundaries are known to introduce recombination-active defect states and electrostatic potential barriers that can modify carrier transport and, in some cases, act as electrical shunts that increase the dark current.[28, 29] Consistent with this behaviour, early studies on thin-film $Zn_3P_2$ have reported that surface states and grain boundaries can strongly influence the photoconductivity response and carrier dynamics.[37]

Previously reported MBE growth of $Zn_3P_2$ on graphene yields grain sizes of up to approximately 2 $\mu$m. In contrast, MOVPE produces grains approaching 10 $\mu$m, which does not represent the maximum size, as full coalescence has not yet been achieved.[27, 41] This substantial increase in grain size indicates a significantly lower nucleation density during MOVPE growth. Such behaviour is consistent with observations in other van der Waals epitaxy systems, such as transition-metal dichalcogenides, where higher growth temperatures and different growth kinetics result in lower nucleation densities and, consequently, larger crystal grains.[42, 43]

The reduced nucleation density observed in MOVPE growth compared to MBE can be attributed to differences in growth kinetics. MOVPE operates at a growth temperature approximately 150 °C higher and involves precursor decomposition at the substrate surface, thereby promoting enhanced adatom mobility and changes in the transport and reactivity of precursor-derived species prior to incorporation.

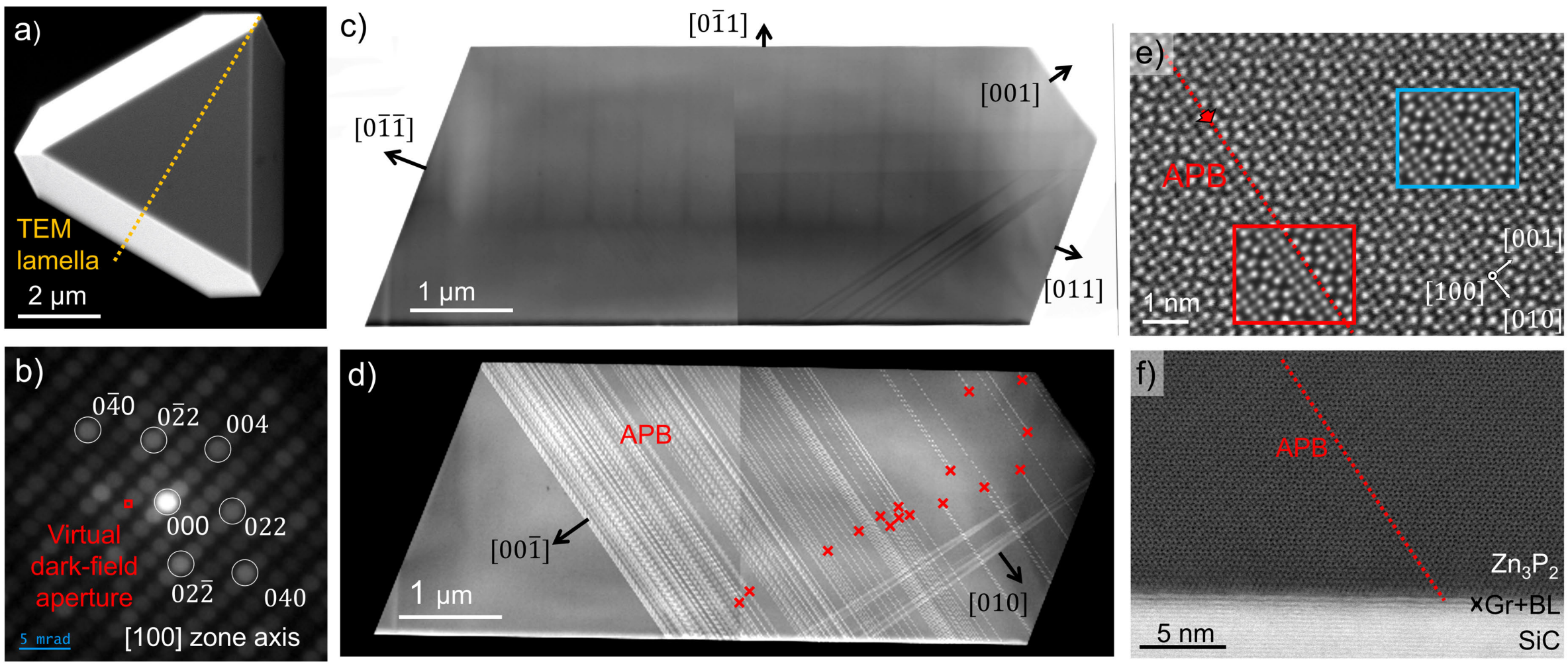

Figure 2: (a) SEM image of a $Zn_3P_2$ crystal indicating the location of the TEM lamella extraction. (b-d) Precession 4D-STEM results. (b) is the electron diffraction pattern observed in the Zn3P2 grain revealing the tetragonal crystal observed in [100] zone axis with out-of-plane orientation [011]. The red square indicates the virtual aperture used to generate the VDF image. (c) and (d) are the VBF and VDF images, respectively, and specific crystal orientations are annotated. The VDF detector was adjusted to generate a high contrast from the APBs. Most of these planar defects run across the grain thickness, while red crosses indicate APB termination. (e) Atomic-resolution HAADF STEM image of Zn3P2, where an APB on (001) is observed along the red dashed line with a displacement vector 1/2[010] indicated with the arrow. Red and blue insets are the image simulation with and without APB, respectively. (f) Atomic-resolution BF-STEM image of the interface between Zn3P2, graphene, buffer layer and SiC. An APB is indicated by the red dashed line.

In contrast, MBE is a ballistic growth method in which species arrive with limited surface diffusion.[44] Since surface diffusion follows an Arrhenius-type behaviour, even moderate increases in temperature lead to an exponential increase in diffusion length.[45] The resulting enhanced mobility reduces the probability of new nucleation events and instead favours the lateral expansion of existing islands. Consequently, fewer nuclei form, provided that the supersaturation remains comparable, allowing grains to grow to significantly larger sizes. Reducing precursor supply could, in principle, achieve a similar effect, but is constrained by the already low growth rates of $Zn_3P_2$ in MOVPE. Although a complete understanding of the influence of the individual growth parameters requires further systematic investigation, the present results demonstrate that MOVPE provides a promising route to substantially increase the grain size of $Zn_3P_2$ grown on graphene, while graphene remains stable under the pre-growth annealing conditions required to achieve these improvements.

### 2.2 Structural defects in MOVPE $Zn_3P_2$

Aberration-corrected scanning transmission electron microscopy (STEM) was performed on a single large $Zn_3P_2$ grain grown at 375 °C and a precursor flux ratio V/II = 41.7, as shown in Fig. 2. Precession 4D-STEM was used on the grain cross-section observed along the [100] orientation, as indicated in the SEM image in Fig. 2a, for large field-of-view analysis of the crystal structure. This technique collects the diffraction patterns at each pixel of the map, allowing to confirm the single-crystalline nature of the grain. A typical diffraction pattern is shown in Fig. 2b.

The virtual bright-field (VBF) image (Fig. 2c) is built using the intensity of the transmitted beam, and the facets are indexed using the reflections observed in the diffraction pattern. The diffraction pattern indicates that the facets correspond to planes with surface normals along the ⟨011⟩ family, i.e., {011} facets. This is in agreement with the literature, where it has been shown that ⟨011⟩ facets are the lowest energy and most stable facets for $Zn_3P_2$.[22, 46]

A virtual dark-field (VDF) image (Fig. 2d), built from the intensity circled by the virtual aperture indicated in the diffraction pattern (Fig. 2b), reveals a network of extended planar defects running across the crystal.[47] These defects appear as parallel lines with well-defined orientations and exhibit a spatially varying density across the lamella. In some regions, the defects are densely packed and continuous over micrometre length scales, while in others they are sparsely distributed and locally interrupted. Such discontinuities are highlighted by red crosses in Fig. 2d.

Atomic resolution high-angle annular dark-field (HAADF) STEM imaging (Fig. 2e) identifies these planar defects as antiphase boundaries (APBs). An APB is defined as a planar defect separating two regions of identical crystallographic orientation that are shifted with respect to each other by a translation vector of the ordered lattice, resulting in a phase shift of the atomic sublattice.[48] In contrast to stacking faults, which arise from a local disruption of the stacking sequence, APBs preserve the lattice orientation but alter the sublattice registry across the boundary. In the present case, the APBs are characterised by a lattice translation along (001) with a displacement vector $R = \frac{1}{2}[110]$ or $R = \frac{1}{2}[010]$.

STEM image simulations (Dr. Probe[49]) superimposed on the experimental images show excellent agreement with both the pristine lattice (blue inset) and the APB-containing lattice (red inset). A second family of planar defects is observed in a different crystallographic orientation, most likely associated with the (110) plane family (Fig. 2d). These defects are also identified as APBs, with displacement vectors $R = \frac{1}{2}[011]$ or $R = \frac{1}{2}[111]$. The (001) APBs continue across intersections with the (110) APBs, undergoing a shift corresponding to the displacement vector of the intersecting boundary.

While the APBs observed here do not exhibit obvious wrong bonds in the idealised structure, local irregularities or terminations of the APBs within the grain, as highlighted in Fig. 2d, may introduce dangling bonds. Additionally, subtle variations in local bonding geometry or coordination at the boundary cannot be excluded. Extended planar defects, including antiphase boundaries, are known to influence the electronic and optical properties of semiconductors and are therefore relevant for the optoelectronic behaviour discussed in the following section.[45] Additional TEM analysis of these defects is provided in Figs. S5-S7.

STEM energy dispersive X-ray (EDX) analysis indicates a homogeneous and stoichiometric composition across the grain. No significant compositional variation is observed across antiphase boundaries (see Fig. S8), indicating that large-scale compositional variation at these defects is absent within the resolution of the technique. However, more subtle atomic-scale variations cannot be excluded.

Strain analysis of the precession 4D-STEM data[50] reveals no measurable strain (below $\pm 0.1\%$) across the $Zn_3P_2$ grain (see Fig. S9), suggesting an effectively strain-relaxed lattice within the investigated region, consistent with growth on graphene via quasi–van der Waals epitaxy.

An atomic resolution image of the $Zn_3P_2$/graphene/SiC interface (Fig. 2f) shows a sharp and well-defined interface without significant roughness in the investigated region. This suggests that the observed planar defects are not directly associated with interfacial roughness; however, given the limited field of view, the formation of the observed antiphase boundaries at other locations or a contribution from the substrate cannot be excluded.

Overall, a dense network of antiphase boundaries with varying density and multiple orientations is observed within a single monocrystalline grain. While many APBs extend across the full thickness of the crystal, others exhibit local irregularities or terminations within the grain.

### 2.3 Cathodoluminescence emission trends

To illustrate the optical properties of MOVPE-grown $Zn_3P_2$, the same grain analysed by STEM in Fig. 2 was investigated prior to lamella preparation using hyperspectral cathodoluminescence (CL) under varying excitation current and temperature conditions.

A representative SEM image of the mapped grain is shown in Fig. 3a, together with spatial maps of the fitted emission peak area (b) and peak centre energy (c), obtained from a 100 × 100 hyperspectral CL map acquired at 10 K (5 keV, 0.5 nA). The luminescence is dominated by a single broad sub-bandgap emission with a pronounced low-energy tail, which was fitted using an exponentially modified Gaussian (EMG) function. The resulting maps therefore represent the spatial distribution of the peak centre energy and integrated emission intensity.

The ROI-Line region (red) exhibits higher emission intensity and an overall blueshift compared to the ROI-Tip region, coinciding with the highest APB density identified in the TEM analysis. In both regions, emission intensity and peak energy show spatial fluctuations across the APBs, while remaining comparatively uniform along directions parallel to them.

A more detailed correlation between APB density and the optical response is provided in Fig. S10, where line-profile analysis and spatial overlays demonstrate that variations in both emission intensity and peak energy closely follow the APB distribution. While the correspondence is not strictly one-to-one, the overall trends indicate a strong, albeit not perfect, relationship between APBs and local optical variations.

As discussed in the structural analysis, the APBs correspond to a crystallographic lattice translation that preserves the overall crystal orientation while altering the sublattice registry. Despite the absence of a pronounced structural disruption, they are associated with clear variations in the local emission. In compound semiconductors, planar defects such as antiphase boundaries can induce significant electronic effects, particularly in polar materials where local phase transitions and associated polarisation fields may lead to carrier localisation and distinct luminescence features.[51–54] Such mechanisms are not expected to apply in the present case, as $Zn_3P_2$ is non-polar and no phase transformation is observed.

Instead, the observed optical variations are likely related to more subtle local perturbations at the defect, such as small variations in bonding geometry or defect irregularities. These can give rise to local potential fluctuations and modified recombination pathways, which may influence the emission intensity and peak energy. A more detailed investigation of the origin of these effects is the subject of ongoing work.[55]

While the spatial shift in peak energy is consistent with modulation of the local electronic potential, variations in emission intensity may additionally reflect differences in recombination pathways. Intensity variations may arise from enhanced non-radiative recombination at terminations within the grain or local irregularities associated with the antiphase boundary, as discussed above. Additional contributions from surface recombination cannot be excluded. Atomic force microscopy and contrast-enhanced SEM images reveal regions with distinct surface textures, likely associated with the high density of APBs, which may locally modify recombination rates and contribute to the observed intensity variations (see Fig. S11 and S12). Low-temperature measurements further indicate that edges between {101} facets act as additional recombination sites due to their rough topography (Fig. S13). The presence of such facets is consistent with a competition between kinetic and thermodynamic growth processes under non-equilibrium conditions, which can promote higher-energy facets alongside the stable {101} planes.[56] As surface recombination is known to play a significant role in $Zn_3P_2$ but can be mitigated through surface passivation,[24, 37, 57] implementing such strategies may provide a pathway to further improve the optical properties of $Zn_3P_2$ grown on graphene.

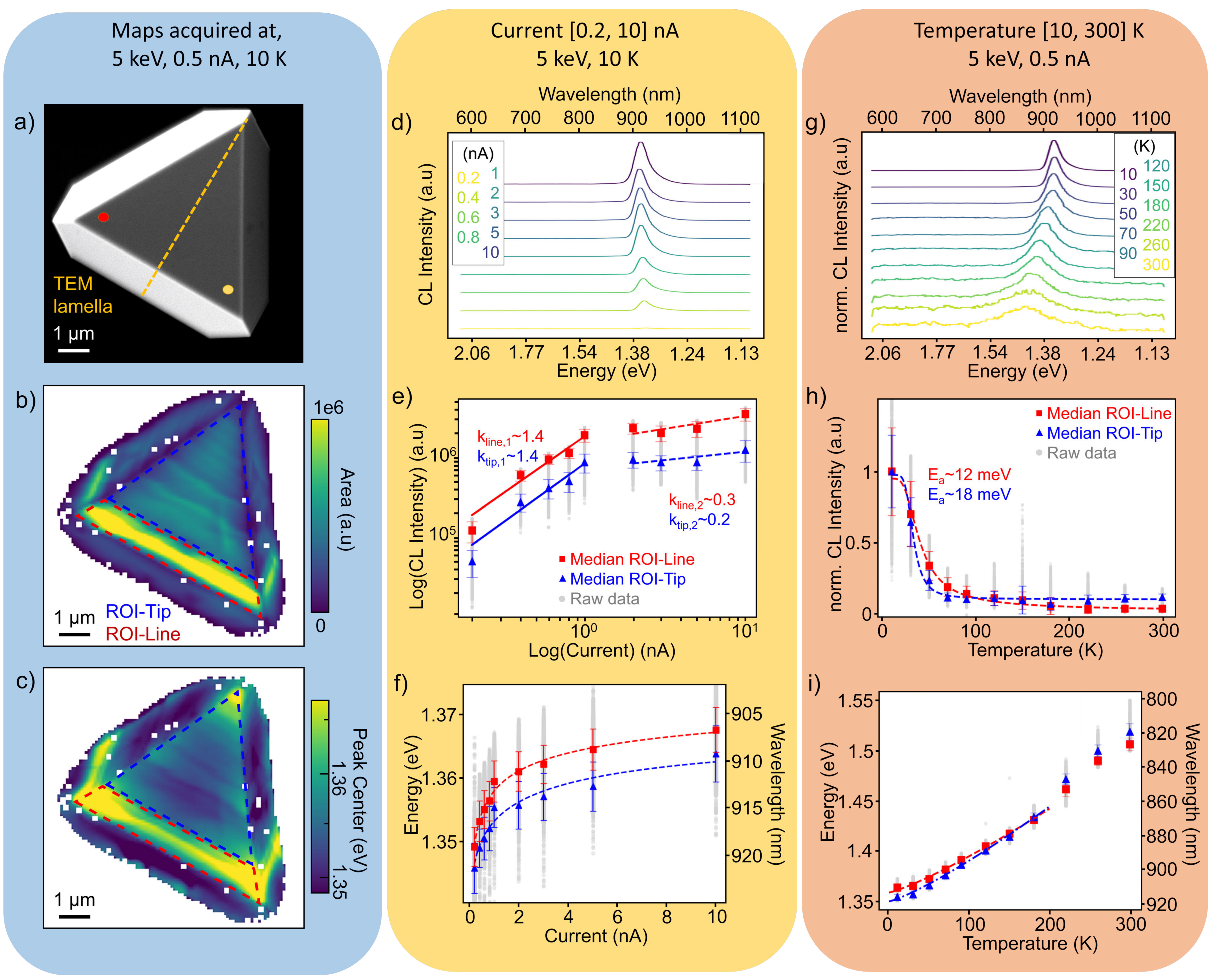

Figure 3: Cathodoluminescence (CL) analysis of MOVPE-grown $Zn_3P_2$, showing the spatial distribution and condition-dependent behaviour of the broad near-infrared emission. Left panel (blue): Hyperspectral CL map (100×100 pixels) acquired at 10 K with a 5 keV electron beam and 0.5 nA current. (a) SEM image of the mapped grain, corresponding to the same grain analysed by TEM in Fig. 2a with indicated TEM lamella position. (b) Map of the fitted emission peak area. (c) Map of the fitted peak centre energy. Two regions of interest (ROIs) are defined: a line-like region (ROI-Line, red dashed) and a tip region (ROI-Tip, blue dashed). Middle panel (yellow): Current-dependent CL measurements (0.2–10 nA) at 10 K. (d) Example spectra from a single pixel set at the yellow-marked spot in (a). (e) Median emission peak area as a function of beam current for both ROIs with power-law fits ($I \propto P^k$). (f) Corresponding evolution of the fitted peak centre energy. Right panel (orange): Temperature-dependent CL measurements (10–300 K) at 0.5 nA. (g) Example temperature series from a single pixel set at the red-marked spot in (a). (h) Median emission peak area versus temperature with Arrhenius fits. (i) Temperature dependence of the fitted peak centre energy. Error bars represent the median absolute deviation within each ROI; grey points correspond to individual pixel values.

The general spectral evolution with excitation current at 10 K is shown in Fig. 3d and with temperature in Fig. 3g. At low temperature, the spectrum consists of a single broad emission peak with a pronounced low-energy tail. The spectral shape remains largely unchanged with increasing excitation current. With increasing temperature, however, the peak broadens and the high-energy side of the spectrum becomes less distinct. Above approximately 150 K, band-edge emission begins to emerge and partially overlaps with the defect-related peak, altering the apparent spectral shape. Such quenching of the band-edge emission below about 150 K is commonly reported for $Zn_3P_2$.[13, 21, 41]

Broad sub-bandgap emission is widely observed in $Zn_3P_2$ and is typically attributed to recombination involving acceptor states or band-tail–like transitions.[13, 21, 27, 41, 58] Although several acceptor and deep defect states have been predicted for $Zn_3P_2$ in both experimental and theoretical studies, the spectra observed here are dominated by a single broad emission band, suggesting that the emission is governed by one dominant recombination channel. Considering the room-temperature bandgap of $Zn_3P_2$ (∼1.5 eV), the observed emission energies are consistent with recombination involving relatively shallow acceptor states or localised band-tail states located on the order of 100–200 meV above the valence band.

The excitation-current dependence of the CL intensity follows a power-law relation $I \propto P^k$, where $I$ is the emission intensity, $P$ the excitation power, and $k$ the exponent associated with the dominant recombination process.[59–62] Both ROIs exhibit a similar behaviour: at low excitation current, the intensity increases with $k \approx 1.4$, followed by a strongly sub-linear regime with $k \approx 0.2$ at higher excitation current (Fig. 3e). A slope around $k \sim 1.4$ is often associated with defect-related transitions such as acceptor-to-band or band-tail recombination, while the sub-linear regime at higher excitation reflects partial saturation of the involved states.[35, 62–66]

Simultaneously, the emission peak exhibits a progressive blueshift with increasing excitation current in both ROIs (Fig. 3f), eventually approaching saturation at higher excitation densities. Such behaviour is widely reported in defect-rich semiconductors and is typically attributed to the filling of localised states or the screening of potential fluctuations.[35, 63–65]

The temperature dependence of the emission intensity was analysed using an Arrhenius model (Fig. 3h). The extracted activation energies are 12 meV for the red ROI-Line region and 18 meV for the blue ROI-Tip region. Although the two regions show slightly different activation energies, the difference is small and does not indicate distinct recombination channels. Instead, the activation energy likely reflects the effective barrier for thermally activated escape from the radiative recombination pathway rather than the binding energy of the radiative state itself. In systems dominated by localised or band-tail states, carriers may thermally redistribute between nearby localised states or escape toward non-radiative recombination centres, resulting in relatively small apparent activation energies.[45] The extracted values (≈10–20 meV) are therefore significantly smaller than the binding energies of shallow acceptors predicted for $Zn_3P_2$, which are typically reported to lie above 100 meV above the valence band.[13, 24, 25, 67]

The peak centre energy increases with temperature for both ROIs (Fig. 3i). Such blueshifting behaviour has been widely reported for recombination involving localised or band-tail states, where carriers thermally escape from deeper localised states and recombine from progressively shallower states as temperature increases.[35, 63–66]

Taken together, the similar excitation-induced blueshift and saturation behaviour, the comparable temperature-dependent trends, and the identical linewidth behaviour across both regions (see Fig. S14), together with the temperature-dependent homogenisation of the emission intensity (see Fig. S15), indicate a common recombination mechanism. The spatial variations in emission energy and intensity are therefore most consistent with modulation of the local potential landscape rather than distinct defect transitions. In particular, regions with a higher density of APBs exhibit a systematic blueshift of the emission peak. The correlation between the line-like emission pattern and the antiphase boundary distribution thus suggests that APBs locally modify the electronic environment, thereby influencing carrier localisation and recombination processes.

While the exact nature of the recombination pathway cannot be unambiguously identified, the continuous spatial variation of peak energy and the absence of distinct linewidth differences suggest recombination involving band-tail states arising from potential fluctuations. Donor–acceptor pair (DAP) transitions cannot be excluded; however, the absence of discrete emission features and the observed spatial and excitation-dependent trends suggest that such processes are unlikely to dominate. First-principles calculations indicate that zinc vacancies have among the lowest formation energies in $Zn_3P_2$,[24] making them likely contributors to the observed emission.

Overall, the CL spectra exhibit a single broad emission band, indicating that the optical response is governed by this dominant recombination channel. The recombination is most consistent with a single dominant pathway involving transitions between conduction-band tail states and shallow intrinsic acceptors, as illustrated schematically in Fig. 4. This assignment is supported by the excitation- and temperature-dependent behaviour: increasing excitation density leads to progressive filling of localised states and a shift of the emission towards higher energies, while increasing temperature promotes thermal redistribution of carriers toward shallower states, resulting in the observed blueshift of the emission peak. Such behaviour is characteristic of recombination involving band-tail states. Although additional pathways, such as band-to-band recombination (observed under PL conditions) and defect-assisted non-radiative recombination, may be present, they do not dominate under the CL conditions investigated here.

### 2.4 Grain boundary recombination and carrier diffusion

Having established the recombination behaviour within individual grains, the role of grain boundaries (GBs) as potential non-radiative recombination centres were investigated using spatially resolved cathodoluminescence.

Figure 5a shows two merged $Zn_3P_2$ grains grown at 375 °C and V/II = 41.7, analysed by hyperspectral CL mapping between 10 K and 300 K. The corresponding CL intensity maps at 10 K and 300 K are shown in Fig. 5b and c. As observed previously for individual grains, line-like intensity variations are visible within the grains. However, at all temperatures, a pronounced decrease in emission intensity occurs at the grain boundary (GB), marked by the dashed red line.

The intensity evolution along the line indicated in Fig. 5a is shown in Fig. 5d for all temperatures. A clear minimum in CL intensity is observed at the GB position, indicating strong non-radiative recombination at the boundary. Although the contrast decreases with increasing temperature (Fig. 5f), the CL intensity at the GB remains reduced by at least a factor of three across all temperatures. Similar emission quenching at grain boundaries has been reported in other thin-film photovoltaic materials such as CIGS, CdTe, kesterites and perovskites, where GB recombination can significantly impact device performance.[30–36] The reduction in GB contrast with increasing temperature indicates a temperature-dependent modification of the recombination landscape, suggesting the activation of additional recombination pathways that compete with the GB recombination channel.

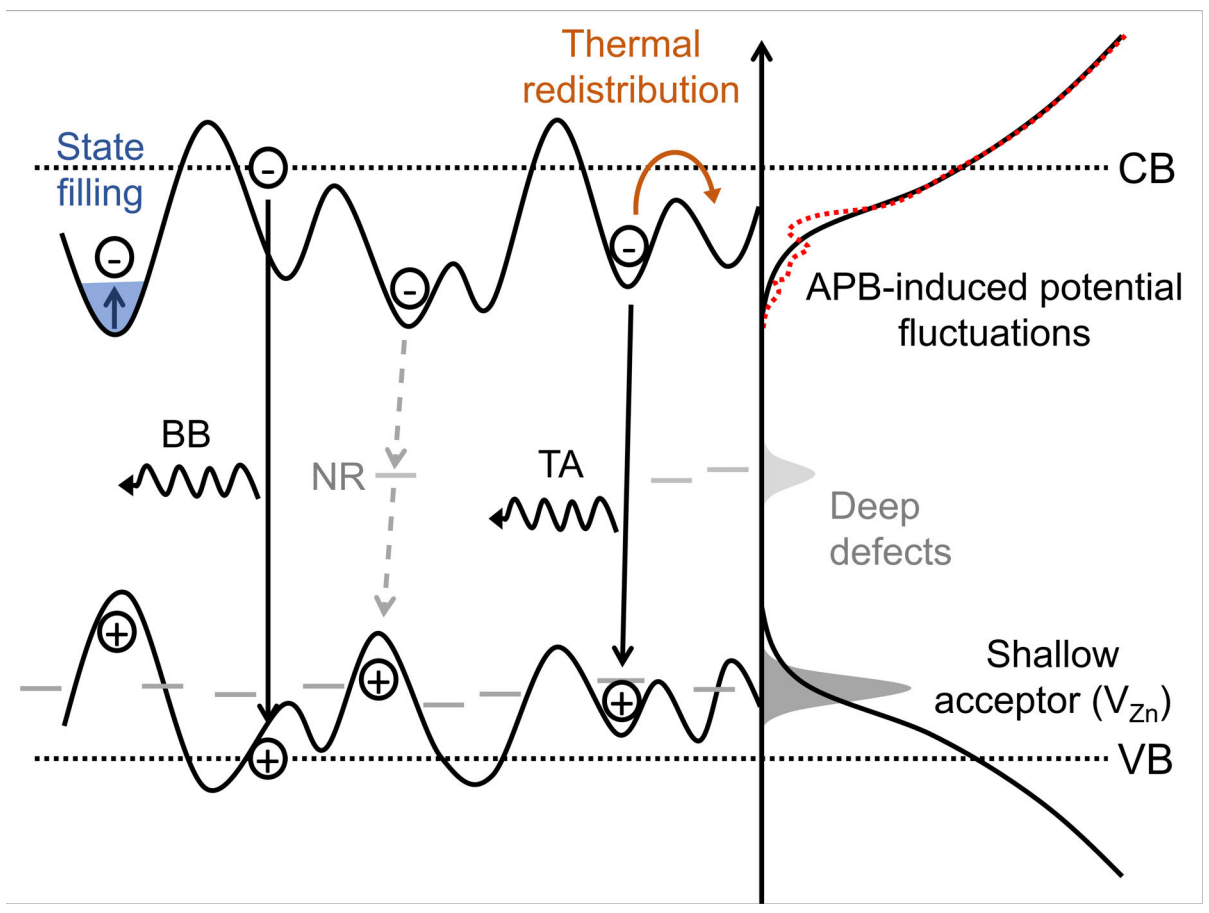

Figure 4: Schematic band diagram illustrating the proposed recombination mechanism in $Zn_3P_2$. The main recombination pathways are shown, including band-to-band (BB) and tail-to-acceptor (TA) radiative transitions, as well as non-radiative recombination (NR) via deep defect states. Increasing excitation density leads to state filling and recombination from higher-energy localised states, while increasing temperature promotes thermal redistribution of carriers toward shallower states, resulting in a blueshift of the emission peak. Antiphase boundaries (APBs) introduce local potential fluctuations that modulate the band-edge landscape.

The spatial recovery of the CL intensity away from the grain boundary is analysed using a steady-state diffusion model. For a strong non-radiative sink at the GB, the carrier density and therefore the CL intensity recover exponentially with distance according to

$$I(x) = I_\infty \left(1 - e^{-x/L}\right)$$

where $I_\infty$ is the intensity far from the GB and $L$ is an effective recovery length. In the strong sink limit, this characteristic length scale follows the same functional dependence as the diffusion length, $L_{\mathrm{diff}} = \sqrt{D\tau_{\mathrm{eff}}}$, as obtained from the steady-state solution of the diffusion equation with recombination,[68] where $D$ is the diffusion coefficient and $\tau_{\mathrm{eff}}$ the effective carrier lifetime.

The recovery behaviour is analysed by plotting the intensity deficit $\Delta I = I_\infty - I(x)$ as a function of distance from the GB (Fig. 5e). Excluding finite-size effects near the grain edges, the intensity deficit follows an approximately exponential recovery. Fitting the near-GB region yields an effective recovery length of approximately 2–3 $\mu$m between 10 K and 300 K (Fig. 5g). This length scale is comparable to reported carrier diffusion lengths $L_{\mathrm{diff}}$ in $Zn_3P_2$, which are on the order of 5–7 $\mu$m.[14, 15] While these literature values correspond to minority carrier diffusion lengths, the present measurements probe ambipolar transport under high-injection conditions; therefore, the extracted length can be used for qualitative comparison of transport length scales, as both quantities are governed by the interplay between carrier diffusion and recombination, but it should not be interpreted as a direct quantitative measure of the minority carrier diffusion length. Reduced effective mobilities have also been reported in $Zn_3P_2$ devices grown via selective area epitaxy, attributed to recombination at mask interfaces, underscoring the sensitivity of transport measurements to the recombination environment.[69]

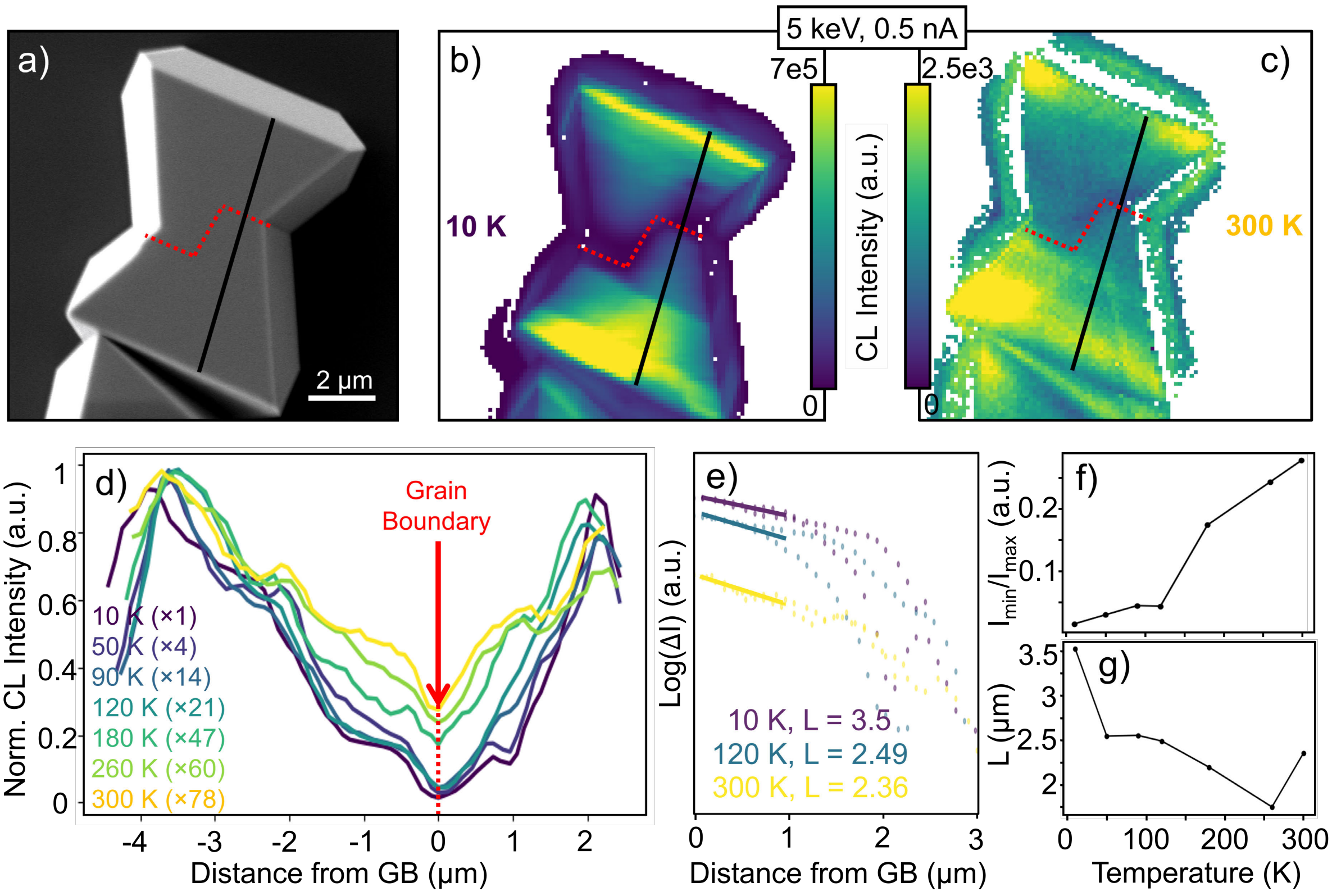

Figure 5: Cathodoluminescence analysis of a grain boundary (GB) between two merged $Zn_3P_2$ grains grown by MOVPE. Hyperspectral CL maps were acquired at 5 keV and 0.5 nA. (a) SEM image of the mapped region with the line scan indicated. (b,c) Integrated CL intensity maps at 10 K and 300 K, respectively, showing strong emission quenching at the grain boundary. (d) Normalised CL intensity profiles across the GB for different temperatures. (e) Recovery of the CL intensity away from the GB plotted as $\Delta I = I_\infty - I(x)$ for selected temperatures, used to extract the effective diffusion length. (f) Ratio of minimum to maximum CL intensity across the GB as a function of temperature. (g) Extracted recovery length $L$ as a function of temperature.

In the present analysis, the grain boundary is treated as an ideal non-radiative sink, such that the recovery profile is described by a single characteristic length scale $L$. While more elaborate models can, in principle, extract the grain-boundary recombination velocity by analysing both the recovery slope and amplitude,[70] this requires independent knowledge of the bulk carrier lifetime and transport parameters, which are not available for the present samples. The analysis is therefore restricted to the effective recovery length $L$. Future time-resolved cathodoluminescence measurements could provide direct access to the carrier lifetime and thereby enable a more quantitative determination of the grain-boundary recombination velocity.

Interestingly, the extracted recovery length remains approximately constant or slightly decreases with increasing temperature (Fig. 5g). This behaviour can be rationalised within the relation for $L$. In semiconductors, the effective carrier lifetime $\tau_{\text{eff}}$ typically decreases with increasing temperature due to enhanced phonon-assisted and defect-mediated non-radiative recombination.[68] To maintain an approximately constant $L$, this reduction in $\tau_{\text{eff}}$ must be compensated by an increase in the diffusion coefficient $D$. Such an increase can arise from enhanced carrier transport at elevated temperatures, for instance due to thermally activated delocalisation of carriers from localised or band-tail states, consistent with the temperature-dependent transport behaviour reported for polycrystalline $Zn_3P_2$ thin films.[41] This interplay provides a consistent explanation for the reduced GB contrast observed at elevated temperatures (Fig. 5f), as increased carrier mobility promotes a more spatially extended redistribution of carriers, while the reduced lifetime enhances non-radiative recombination, in agreement with the activation of additional recombination pathways discussed above.

Grain boundaries (GBs) in $Zn_3P_2$ have previously been identified as charged, recombination-active regions that can dominate carrier transport.[37] In agreement with this picture, the pronounced CL quenching observed at GBs demonstrates that they act as efficient non-radiative recombination centres. The extracted recovery length, which approaches a significant fraction of the grain size, indicates that carriers diffuse toward these non-radiative recombination regions, leading to a spatial reduction in carrier density near the GBs. The presence of grain-boundary charge and associated electrostatic barriers further modifies the local carrier distribution. Since the quasi-Fermi level splitting is directly governed by the local electron and hole populations, this results in a corresponding spatial variation and reduction of the quasi-Fermi level splitting in the vicinity of GBs.[68, 71]

Although some earlier studies reported a limited impact of GBs on device performance,[16] these observations can be reconciled with the present results by considering differences in microstructure. In particular, those studies were conducted on materials with substantially larger grain sizes, where the lower grain-boundary density likely mitigated their overall impact. In the present case, the combination of strong non-radiative recombination and a recovery length comparable to the grain size indicates that GBs can influence carrier transport over a large fraction of the absorber volume. Similar recovery lengths have been observed from room-temperature photoluminescence measurements, as shown in Fig. S16.

These observations highlight the importance of controlling grain-boundary density and activity in $Zn_3P_2$. Minimising grain boundary density through larger grain growth, therefore, represents an important pathway toward improving the optoelectronic quality of $Zn_3P_2$ absorbers.

### 2.5 MBE vs MOVPE thin-film quality

The analyses above reveal how defects within grains and grain boundaries shape the local optoelectronic behaviour of $Zn_3P_2$. In the following, these insights are used to compare the properties of thin films grown by MBE and MOVPE, highlighting how differences in growth conditions influence defect-related sub-bandgap emission and grain structure.

Figure 6 compares the CL emission at 10 K of coalesced $Zn_3P_2$ thin films grown by MBE and MOVPE under markedly different conditions. The MBE sample was grown at a nominal temperature of 230 °C with a V/II ratio of 3.4, while the MOVPE sample was grown at 375 °C with a V/II ratio of 41.7. In both cases, the film thickness exceeds the electron interaction volume under the CL conditions, ensuring that the measured emission is representative of the bulk film. The samples are therefore representative of the respective growth approaches: MBE films show limited variation across the explored growth window, while the MOVPE sample corresponds to the only fully coalesced thin film obtained under the investigated conditions. It should be noted that the V/II values are not directly comparable due to differences in precursor chemistry and growth dynamics between the two techniques.

Measurements on the MBE-grown film were performed on an exfoliated flake detached from the graphene substrate, similar to previously reported measurements.[27] The MOVPE-grown film, in contrast, was measured while still in contact with the graphene layer. Together with possible differences in film thickness, these factors influence the detected CL intensity, as contact with conductive substrates such as graphene can lead to partial luminescence quenching due to carrier extraction or interface recombination.[27] Consequently, the absolute CL intensities between the two samples cannot be directly compared, and the following discussion focuses primarily on spatial emission variations and spectral features.

Two main emission bands can be identified in the MBE-grown film. The first, centred around 1.37 eV, is referred to as the high-energy (HE) emission, while a second band appears below approximately 1.34 eV and is denoted as the low-energy (LE) emission. The waterfall plot of spectra extracted along the line shown in Fig. 6a (Fig. 6c) reveals strong spatial variations in both peak position and relative intensity for the HE and LE emissions, indicating significant local variations in the optical properties of the film, with the LE emission being highly localised. Similar spatially localised low-energy emission has been reported previously in MBE-grown $Zn_3P_2$ thin films on graphene.[27]

In contrast, the MOVPE-grown thin film exhibits only the emission previously discussed in Fig. 3, centred around 1.35 eV. As shown in the spectral evolution along the line scan in Fig. 6d, spatial variations in intensity and peak position are still observed, likely related to APBs and grain boundaries, but no additional low-energy emission band is present.

Despite differences in peak position and linewidth between the HE emissions of the MBE- and MOVPE-grown films, their similar temperature- and current-behaviour suggests that they originate from the same underlying recombination mechanism (see Fig. S17 and S18 for temperature- and current-behaviour of MBE-grown thin films). The observed spectral shift and broadening are likely related to differences in band-tail states, composition, or strain arising from the distinct growth conditions.

The strong spatial localisation of the LE emission suggests that it originates from local variations in the electronic structure. Previous studies on comparable $Zn_3P_2$ thin films have reported the presence of highly localised strain fields,[27], which could give rise to such emission features. Density functional theory calculations (Fig. S19) further show that strain can induce a significant reduction of the bandgap, supporting the interpretation that the LE emission originates from strain-modified band-tail states.

Comparison of the SEM image and the integrated CL map for the MOVPE-grown film (Fig. 6e,f) reveals a clear correlation between morphology and emission intensity, with reduced CL intensity occurring predominantly at grain boundaries and grain edges. This behaviour contrasts with the MBE-grown film, where strong emission variations occur within the grains.

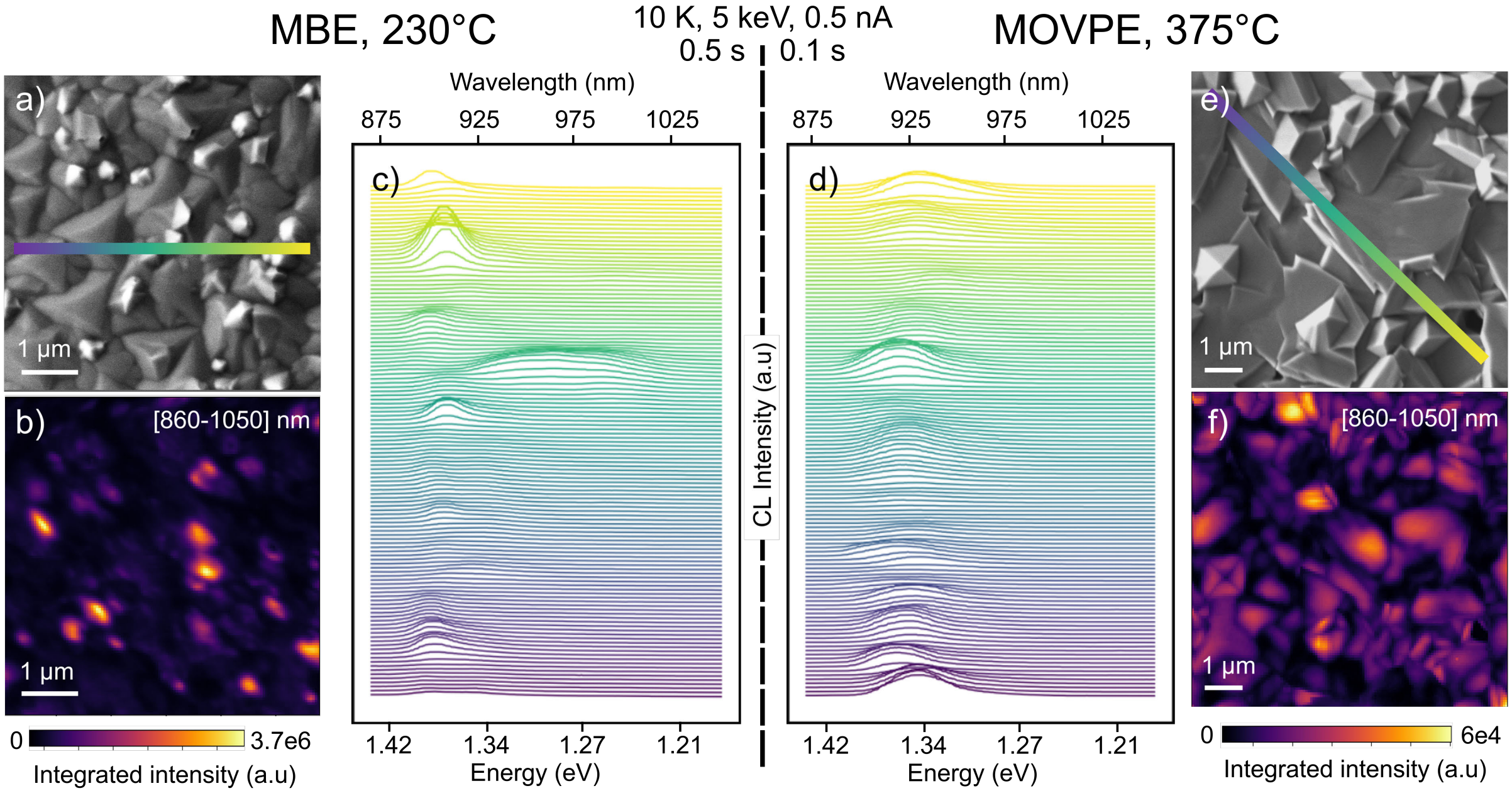

Figure 6: Comparison of cathodoluminescence of $Zn_3P_2$ thin films grown by MBE (left) and MOVPE (right) at 10 K. (a) SEM image of the MBE sample. (b) Integrated CL intensity map in the spectral range 860–1050 nm of the CL map acquired at 10 K for the same region as shown in the SEM image in (a). (c) CL spectra extracted along the line indicated in (a), plotted with an offset for clarity. The colour coding corresponds to the spatial position along the scan direction. (d) CL spectra along the line indicated in (e) for the MOVPE sample, displayed with the same offset and colour scheme as in (c) for direct comparison. (e) SEM image of the second sample with the corresponding line scan indicated. (f) Integrated CL intensity map (860–1050 nm) of the region shown in the SEM image in (e). All CL measurements were performed at 10 K with a 5 keV beam energy and 0.5 nA current. MBE sample pixels were acquired for 0.5 s and MOVPE for 0.1 s.

Overall, the absence of the strain-related low-energy emission in the MOVPE-grown thin film indicates a reduced density of localised strain fields and associated defect states compared to the MBE-grown material. Together with the clear correlation between morphology and emission intensity, this suggests that MOVPE growth produces material with improved optoelectronic quality and fewer non-radiative defects within the grains. This improvement is attributed in part to the higher growth temperatures accessible in MOVPE, which enhance adatom mobility and promote defect relaxation during growth. In contrast, MBE growth of $Zn_3P_2$ is inherently limited to significantly lower substrate temperatures due to the high desorption rates of both Zn and $P_2$,[72] which restricts surface diffusion and favours the incorporation of local strain and structural defects. As a result, the extended adatom diffusion lengths achievable in MOVPE likely play a key role in reducing the formation of localised strain fields and associated band-tail states at grain boundaries.

## 3 Conclusion

In this work, we demonstrate the growth of $Zn_3P_2$ on graphene by metal–organic vapour phase epitaxy (MOVPE) and directly link structural defects to optical emission through correlative transmission electron microscopy and cathodoluminescence spectroscopy. In-situ treatment of the graphene substrate with $HCl_{(g)}$ prior to growth improves the crystal quality of the resulting films, highlighting the importance of surface preparation for achieving high-quality $Zn_3P_2$ layers. Further investigation of the MOVPE growth window may enable improved control of nucleation and the development of more homogeneous films over larger areas.

Transmission electron microscopy reveals that individual grains are monocrystalline and contain a high density of antiphase boundaries. Within the grains, no significant elemental composition or strain is observed, including around the APBs. This behaviour is consistent with quasi–van der Waals epitaxy, where weak interfacial coupling relaxes lattice-matching constraints. Correlative CL measurements performed on the same grain reveal spatial variations in emission intensity and peak energy that coincide with the density of APBs. This behaviour suggests that APBs locally perturb the electronic landscape, likely introducing potential fluctuations that modulate the band-tail states. Temperature- and current-dependent measurements indicate that the dominant sub-bandgap emission originates from a band-tail-to-acceptor transition.

Spatially resolved CL analysis of grain boundaries shows pronounced non-radiative recombination, with an effective carrier diffusion length of approximately 2 $\mu$m. This value is comparable to roughly half the grain size, indicating that grain boundaries can significantly influence carrier transport and recombination dynamics in polycrystalline $Zn_3P_2$.

Finally, extending this analysis to the thin-film level, comparison between MBE- and MOVPE-grown samples reveals clear differences in their optical properties. While MBE-grown films exhibit additional spatially localised low-energy emission attributed to strain-related band-tail states, MOVPE-grown films show only a single emission peak present across the entire film, together with a clear correlation between morphology and emission intensity. The absence of strain-related sub-bandgap emission suggests that MOVPE growth reduces the formation of localised strain fields and associated defect states at grain boundaries.

Overall, MOVPE-grown films exhibit larger grains and the suppression of additional defect-related sub-bandgap emission observed in MBE-grown material, indicating improved optoelectronic material quality compared to MBE growth. At the same time, the clear identification of grain boundaries as dom-inant non-radiative recombination centres highlights the importance of controlling the microstructure of polycrystalline $Zn_3P_2$.

These results highlight that improving grain size and mitigating grain-boundary recombination, for example through optimised growth conditions or targeted passivation strategies, represents a promising pathway toward high-performance earth-abundant $Zn_3P_2$ photovoltaic absorbers.

## Experimental Methods

### $Zn_3P_2$ growth procedure

$Zn_3P_2$ was grown by molecular beam epitaxy (MBE) and metal-organic vapour phase epitaxy (MOVPE).

**MBE growth:** MBE growth was performed in a Veeco GENxplor system. Zn (6N purity) was supplied from a valved cracker cell, while $P_2$ was provided by a valved GaP (6N purity) source (MBE-Komponenten) equipped with an integrated Ga trapping system. The V/II flux ratio ($P_2$/Zn) was determined from beam flux monitor measurements performed immediately after growth.

The substrate temperature was measured using a thermocouple positioned above the manipulator; all reported temperatures are nominal values. Prior to growth, the substrates were annealed under ultra-high vacuum (UHV, below $5.0 \times 10^{-10}$ Torr) at 650 °C for 2 h in a preparation chamber adjacent to the growth module to desorb water and airborne contaminants. An additional 1 h annealing step at 650 °C was performed in the growth chamber.

After annealing, the substrate was cooled to the growth temperature and exposed to $P_2$ for 5 min before initiating $Zn_3P_2$ growth. The beam equivalent pressures (BEP) were $1.4 \times 10^{-6}$ Torr for Zn and $3.4 \times 10^{-7}$ Torr for $P_2$. Growth duration was 390 min.

**MOVPE growth:** MOVPE growth was carried out in an Aixtron 3×2” close-coupled showerhead (CCS) MOCVD reactor operating at 100 mbar with a total flow rate of 8000 sccm hydrogen carrier gas.

Prior to growth, the graphene substrates were degassed at 650 °C for 15 min under a $PH_3$ partial pressure of 0.1 mbar, with or without additional $HCl_{(g)}$ flow of $1.76 \times 10^{-3}$ mbar. The reactor was then cooled to the growth temperature (350 °C or 375 °C). Once thermal stabilisation was reached, $Zn_3P_2$ growth was initiated using diethylzinc (DEZn) as the Zn precursor and $PH_3$ as the P precursor. The DEZn partial pressure was $7.19 \times 10^{-3}$ mbar, while the $PH_3$ partial pressure ranged from $2.25 \times 10^{-1}$ mbar to $3.0 \times 10^{-1}$ mbar. Growth durations ranged from 120 to 210 min.

**Graphene substrates:** $Zn_3P_2$ was deposited on epitaxial monocrystalline graphene grown on the Si-face of 6H-SiC by chemical vapour deposition (CVD) under hydrogen.[73] The graphene was synthesised at 1550 °C under low hydrogen partial pressure,[74] under low hydrogen partial pressure, resulting in the self-limited growth of monocrystalline monolayer graphene with approximately 5% bilayer coverage

A carbon buffer layer composed of $sp^2$ and $sp^3$ hybridised carbon is present at the graphene/SiC interface. The SiC substrates had an average miscut of approximately 0.07°, leading to step heights between 0.25 and 1.5 nm and terrace widths up to 400 nm.

## Measurement techniques

**Scanning electron microscopy (SEM):** SEM images were acquired using a Zeiss Merlin field-emission SEM equipped with a Gemini column or a Zeiss Gemini 500. Imaging was performed using an in-lens secondary electron detector at an acceleration voltage of 2 keV or 3 keV and a beam current of 80 pA.

**Atomic force microscopy (AFM):** AFM measurements were carried out using a Bruker FastScan system operated in tapping mode. Bruker FluidScan+ probes were used for all measurements.

**Photoluminescence spectroscopy (PL):** Room-temperature PL measurements were performed using a Renishaw inVia system with a 532 nm excitation laser and a 100× objective (NA = 0.85), corresponding to an illumination intensity of approximately 393 $\mu$W/$\mu$m$^2$. The laser power was 180 $\mu$W for all measurements. Single-point spectra were acquired with 1 s integration time and averaged over 60 accumulations, whereas mapping measurements were performed with a single acquisition per pixel. A 300 lines mm$^{-1}$ grating was used.

**Raman spectroscopy:** Raman measurements were performed at room temperature using the same Renishaw inVia setup with a 532 nm excitation laser and a 100× objective (NA = 0.85). Spectra were acquired with 5 s integration time and averaged over 60 accumulations. The laser power was approximately 180 $\mu$W. An 1800 lines mm$^{-1}$ grating was used.

**Cathodoluminescence spectroscopy (CL):** Temperature-dependent CL measurements (10–300 K) were conducted using an Attolight Allalin system. Spectra were recorded with an Andor Newton DU920P-BEX2-DD Si-CCD detector at an acceleration voltage of 5 keV, beam currents between 0.2 and 10 nA, and acquisition times ranging from 0.1 to 0.5 s per pixel. A 150 lines mm$^{-1}$ grating with a blaze wavelength of 500 nm was used.

CL spectra were corrected by subtracting the CCD dark count. To reduce high-frequency noise, a Savitzky–Golay low-pass filter (window size = 5 pixels, polynomial order = 3, derivative = 0) was applied.

When fitting was performed, emission peaks were modelled using an exponentially modified Gaussian (EMG) function, which reproduces the asymmetric peak shape observed in the spectra. The extracted fit parameters were analysed using median values rather than arithmetic means to obtain robust representative values. The median was selected because it is less sensitive to skewed distributions and rare high-intensity events that can bias the mean. For each defined region of interest (ROI), the upper 5% of the parameter distribution was excluded prior to calculating the median. This approach reduces the impact of localised bright emission spots or occasional fitting artefacts while preserving the dataset's statistically relevant behaviour.

**Transmission Electron Microscopy (TEM):** TEM samples were prepared using a lift-out technique on a Zeiss CrossBeam 550 dual-beam focused ion beam-SEM (FIB-SEM) instrument. TEM experiments were conducted on a double-corrected FEI Titan Themis microscope at 300 kV. Atomic-resolution scanning transmission microscopy (STEM) images used a convergence angle of 20 mrad, a beam current of 100 pA, and collection angles of >50 mrad and <20 mrad for HAADF and BF images, respectively. For EDX, the beam current was increased to 270 pA, and the characteristic X-rays were collected with silicon drift detectors (super-X system). For the precession 4D-STEM experiment, a NanoMEGAS

Topspin system was used, with data collected using a Quantum Detectors MerlinEM direct-detection camera at a convergence angle of 0.75 mrad, a beam current of 470 pA, a camera length of 230 mm, and a precession angle of 1°. The entire $Zn_3P_2$ grain was mapped by stitching two 4D-STEM maps with a 15 nm pixel size and a 6 ms dwell time. Strain maps were processed using in-house codes developed as a plugin for GMS software, named 4Dview, which appears to be quite robust against artefacts from diffraction contrasts. Multislice atomic-resolution STEM image simulations were performed with Dr Probe software,[49] using a 105 nm-thick crystal sample from the CIF file of P4nmc $Zn_3P_2$, which has been modified to introduce the specific antiphase boundary defect.

**Density Functional theory (DFT):** Ab initio calculations were performed using the Vienna Ab initio Simulation Package (VASP, version 6.3.2).[75, 76] The initial crystal structure of $Zn_3P_2$ was obtained from the Materials Project (mp-2071).[77] Strain was applied via linear deformations in increments of 0.5 % over a range from -3 % to +3 % along three directions orthogonal to the $[\bar{1}11]$ plane. Ionic positions were fully relaxed using the PBEsol exchange-correlation functional,[78] with a plane-wave cutoff energy of 350 eV, until the residual forces on all atoms were below 1 meV/Å. Brillouin zone integrations were performed using a Γ-centered $9 \times 9 \times 6$ k-point mesh. Subsequently, electronic band structures were calculated along high-symmetry paths in reciprocal space using 278 k-points and the modified Becke–Johnson potential.[79]The same cutoff energy of 350 eV was employed, and both spin-orbit coupling and non-spherical contributions to the charge density were included.

## Authors Contribution

TH carried out epitaxial growth (MBE) and optical characterisation together with MH, as well as the writing process. SES and AU carried out the epitaxial growth by MOVPE with input from SL and KAD. GG fabricated the TEM lamella, and VB undertook the STEM characterisation, including STEM-EDS and GPA analysis. AEA and AM produced and provided the Si-face H-CVD graphene substrates. NK conducted the DFT simulations under SB's supervision. RL contributed to the design and realisation of the epitaxial growth (MBE). SES and AFM outlined the research project, directed the work, and contributed to the manuscript writing. AFM also supervised the project. All authors reviewed and agreed to the final manuscript.

## Acknowledgements

Authors from Lund University, EPFL, Ruhr University Bochum (including ICAMS), and Friedrich Schiller University Jena acknowledge support from Horizon Europe through the Pathfinder project SOLARUP (project number: 101046297). Lund University authors further acknowledge support from NanoLund and the Lund Nano Lab (Myfab Lund). EPFL authors acknowledge additional funding from the COST grant (IZCOZ0_220221). Graphene growth at CRHEA was supported by the ANR through the VanaSiC project (ANR-22-CE24-0022-01) and by the ANR in the framework of the PEPR Electronique through the ADICT project (ANR-22-PEEL-0011).

# Supplementary Information for : Cathodoluminescence Analysis of Defects and Grain Boundaries in $Zn_3P_2$ Thin Films Grown on Graphene by MOVPE and MBE

Thomas Hagger[1], Mohammadreza Hassanzadeh[1], Aidas Urbonavicius[2], Ahmed El Alouani[3], Victor Boureau[4], Gulnaz Ganeeva[4], Nico Kawashima[5,6], Raphael Lemerle[1], Kamil Artur Wodzislawski[1], Sebastian Lehmann[2], Kimberly A. Dick[2], Silvana Botti[5,6], Adrien Michon[3], Anna Fontcuberta i Morral[1], and Simon Escobar Steinvall[2]

[1]Laboratory of Semiconductor Materials, Institute of Materials, School of Engineering, École Polytechnique Fédérale de Lausanne (EPFL), 1015 Lausanne, Switzerland

[2]Centre for Analysis and Synthesis and NanoLund, Lund University, Box 124, 221 00 Lund, Sweden

[3]Université Côte d'Azur, CNRS, CRHEA, Rue Bernard Grégory, 06560 Valbonne, France

[4]Interdisciplinary Center for Electron Microscopy (CIME), EPFL, Lausanne, Switzerland

[5]Research Center Future Energy Materials and Systems of the University Alliance Ruhr and ICAMS, Ruhr University Bochum, Universitätsstraße 150, D-44801 Bochum, Germany

[6]Institut für Festkörpertheorie und-optik, Friedrich-Schiller-Universität Jena, Max-Wien-Platz 1, 07743 Jena, Germany

Corresponding authors: anna.fontcuberta-morral@epfl.ch; simon.escobar_steinvall@chem.lu.se

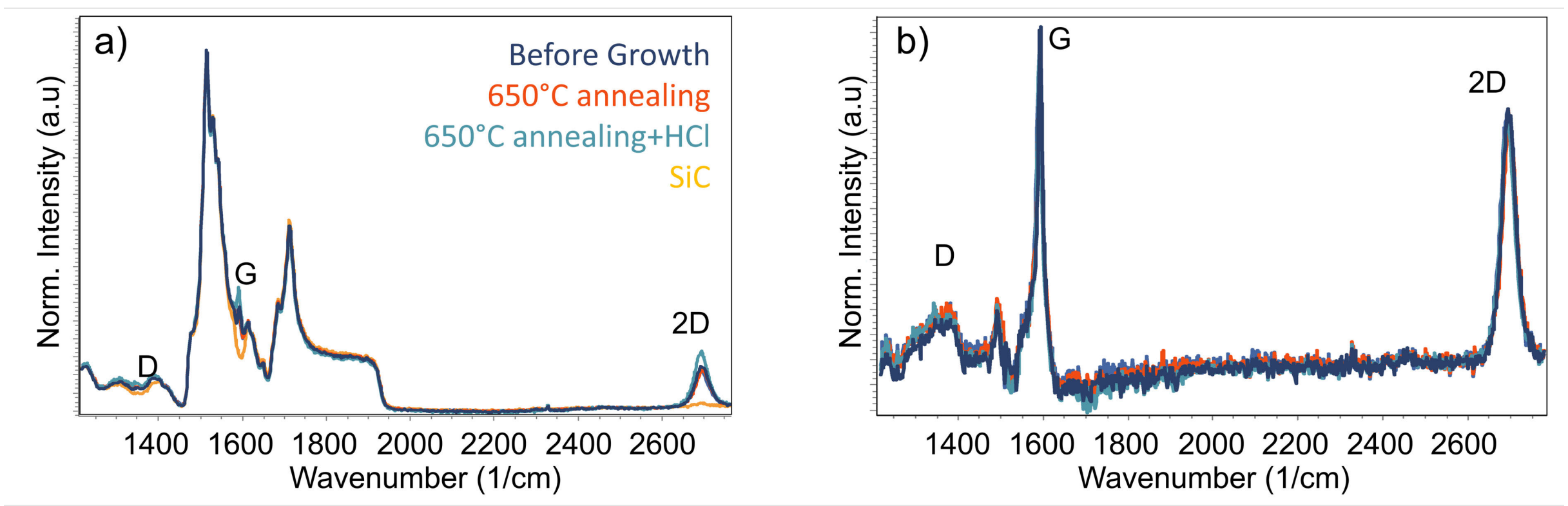

Figure S1: Raman spectra of the graphene substrate before and after the growth of $Zn_3P_2$ by MOVPE. (a) shows the raw spectra and (b) the graphene spectra with subtracted SiC background. No damage to the graphene was induced during any of the processes, as evidenced by the unchanged intensity of the D peak.

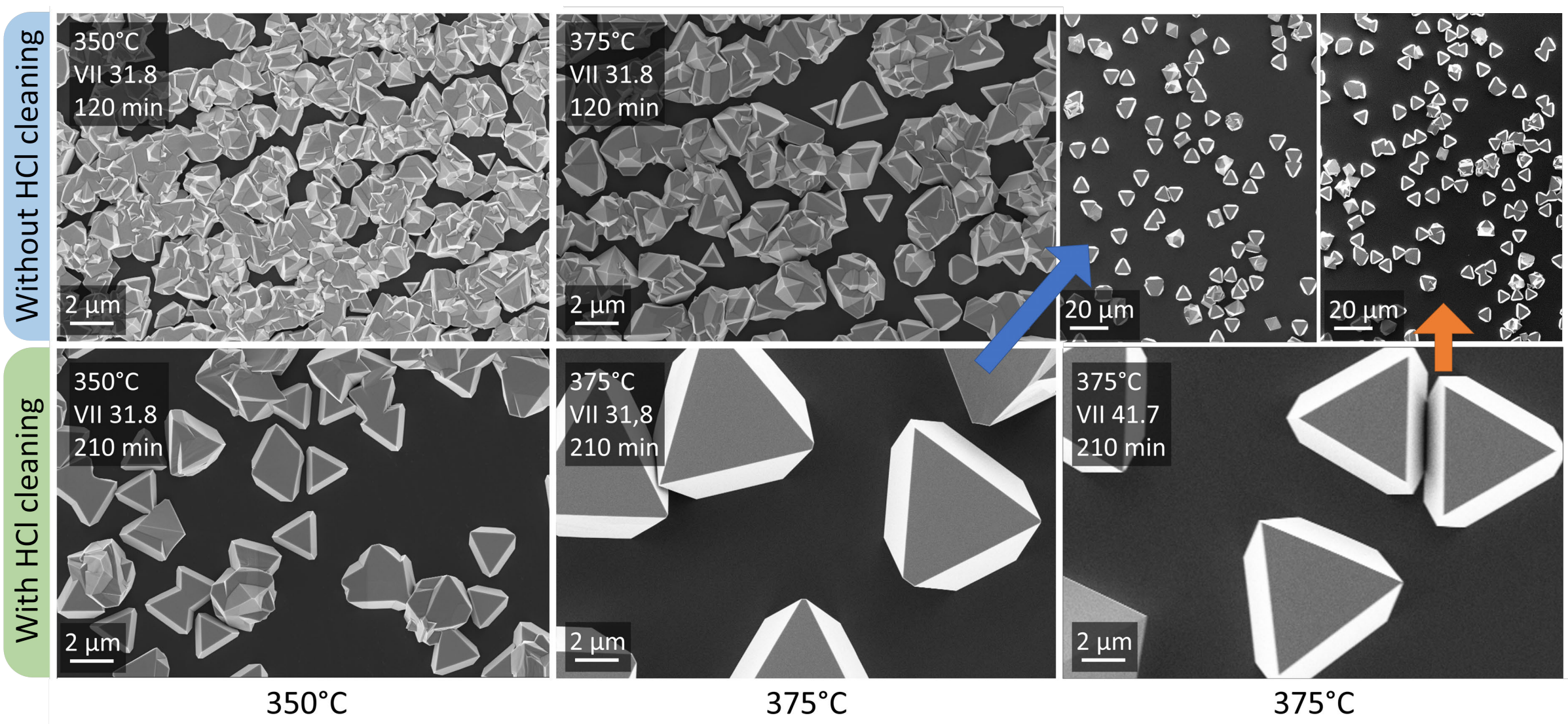

Figure S2: MOVPE growth outcome of $Zn_3P_2$ as presented in the main text with an additional increase in V/II flux ratio. No significant changes in nucleation density are observed by increasing the V/II flux ratio.

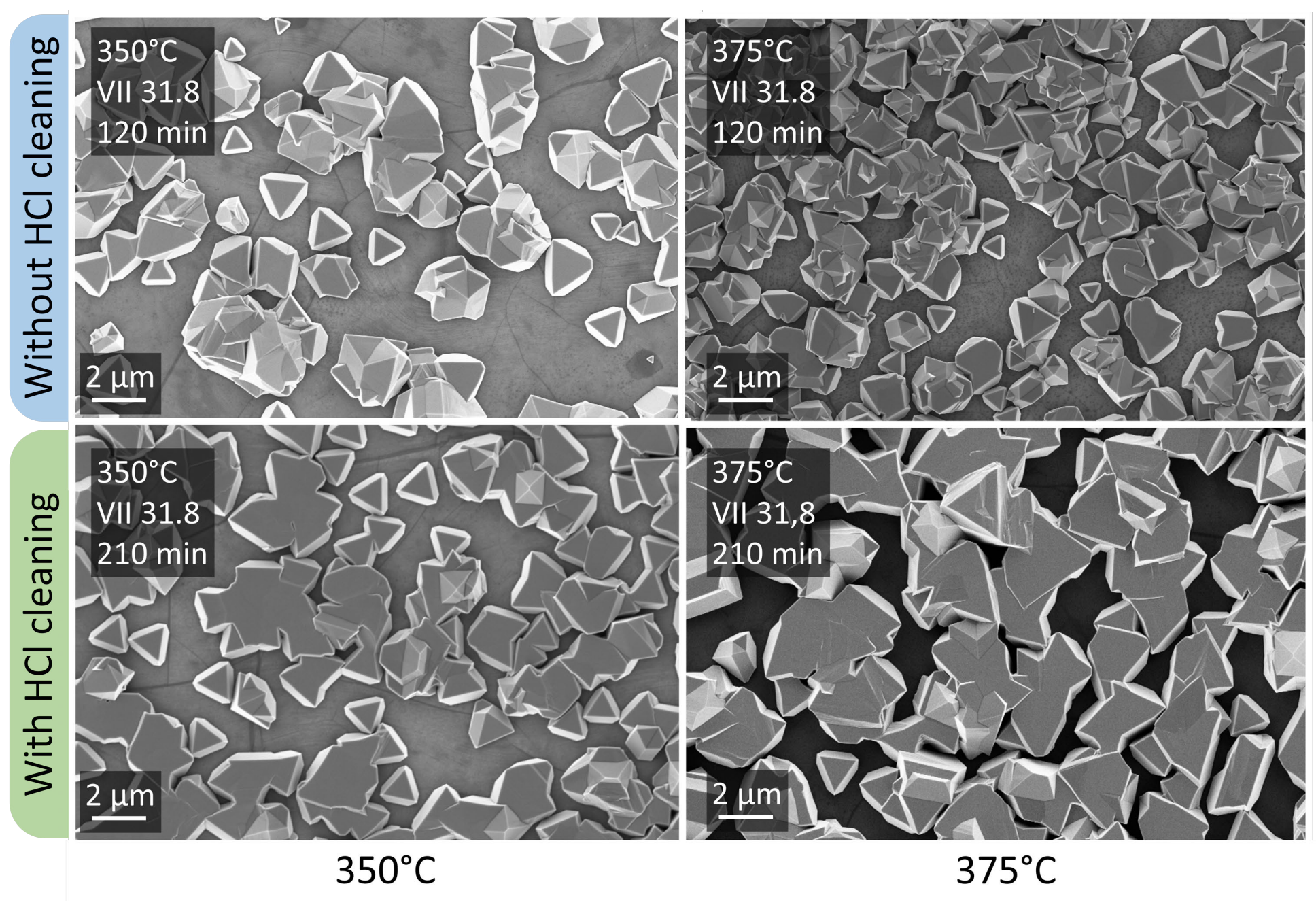

Figure S3: MOVPE growth outcome of $Zn_3P_2$ on the commercially available wet-transferred graphene from Graphenea.

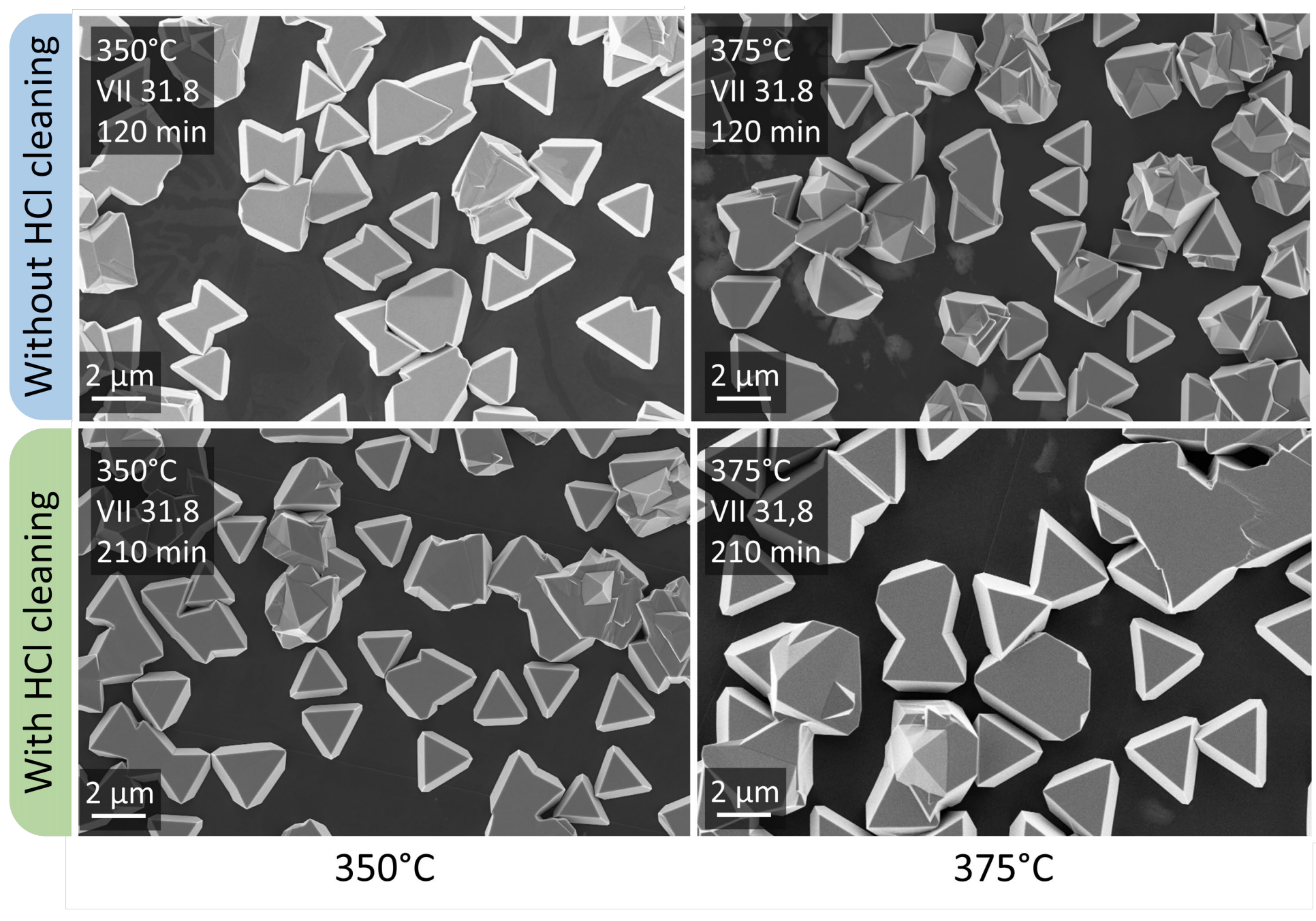

Figure S4: MOVPE growth outcome of $Zn_3P_2$ on the commercially available epitaxial graphene on Si-face SiC obtained via high-temperature annealing from Graphensic.

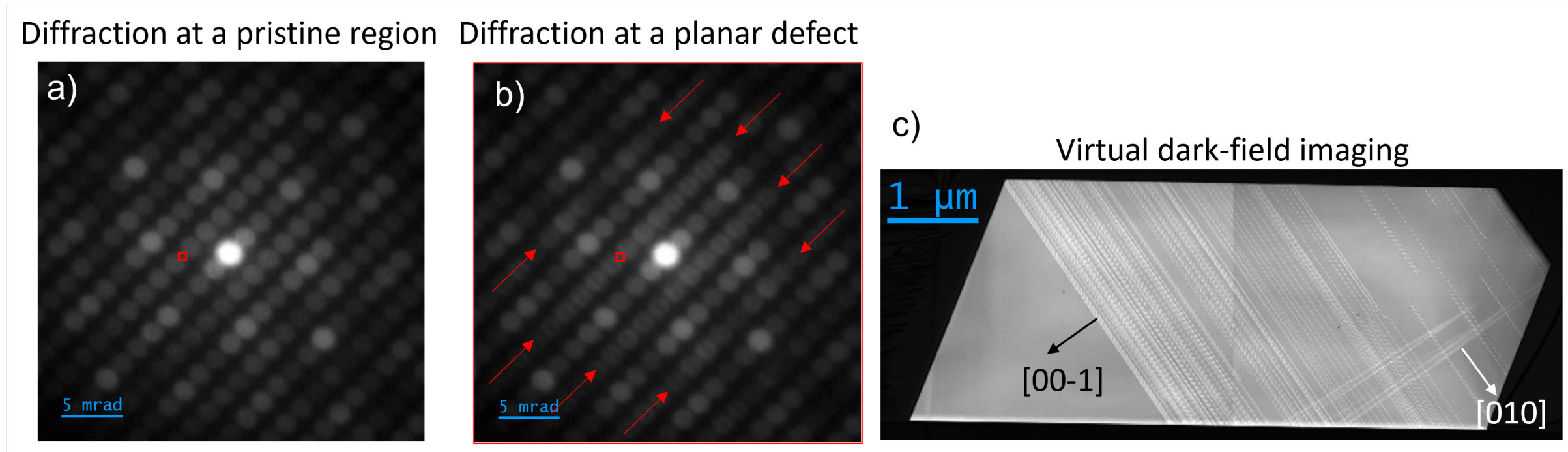

Figure S5: 4D-STEM measurements showing the diffraction patterns for pristine Zn3P2 (a) and Zn3P2 containing an APB (b). The APB perturbs the Bragg intensities of the systematic rows indicated by the red arrows, corresponding to reflections of type (0 j k) with j = 2n + 1 and k an integer. Thus, the virtual aperture drawn as a red square in the diffraction patterns is used to generate high contrast of the APB in the virtual dark-field image shown in (c).

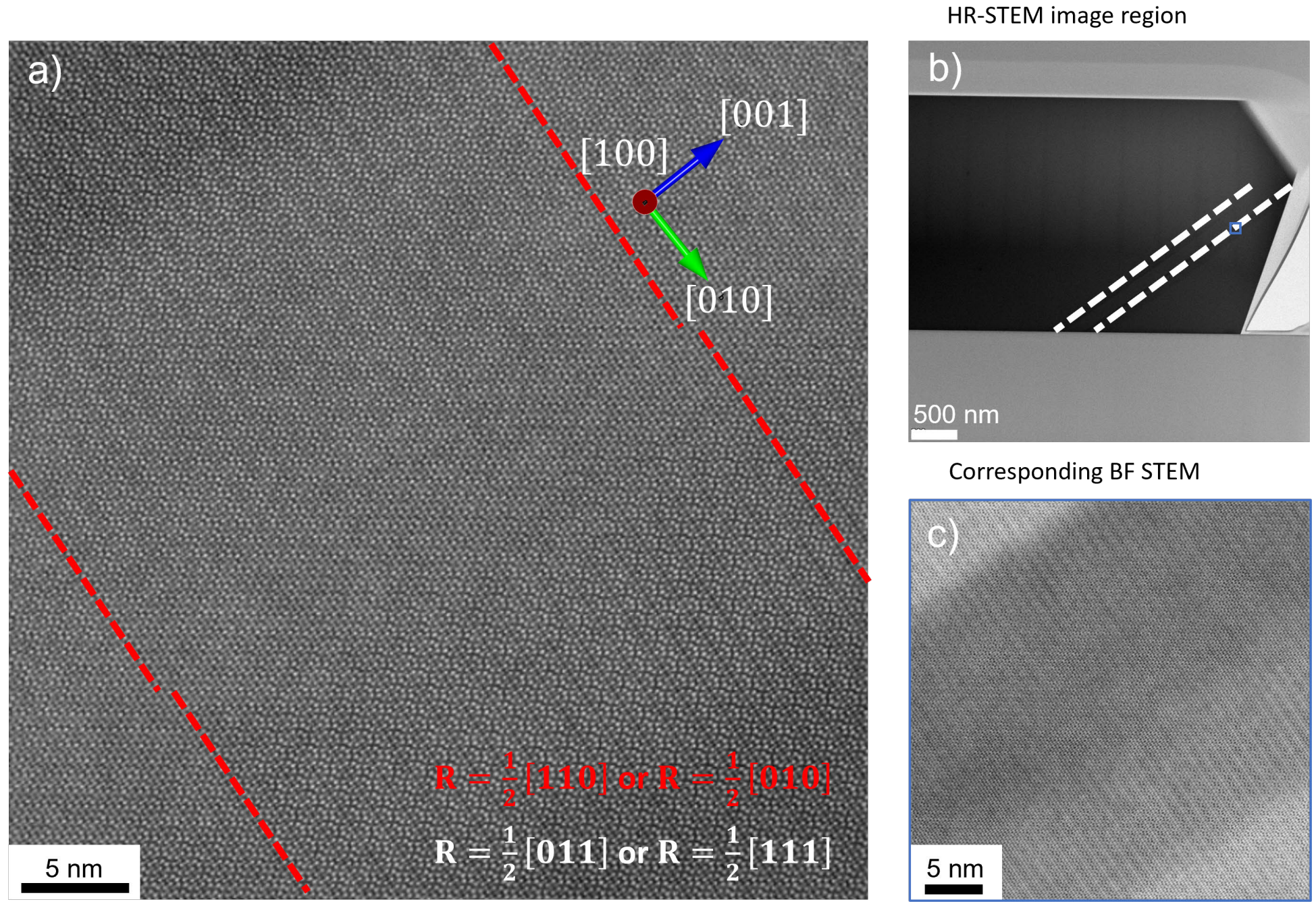

Figure S6: AC-HAADF-STEM image with atomic resolution of the APB with direction <020> visible as a broad line, indicated in white, and the [002] APB indicated in red, which is displaced by the first one.

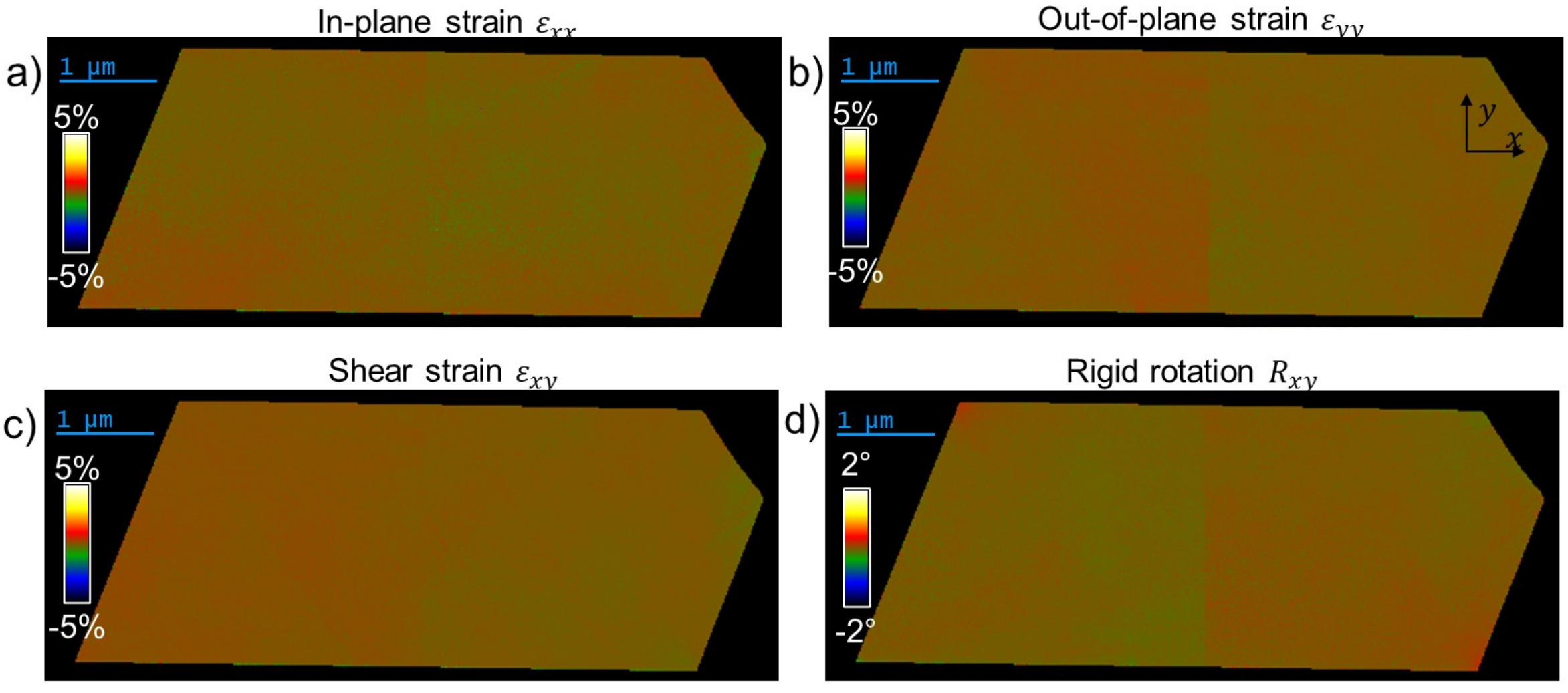

Figure S7: Precession 4D-STEM analysis of strain and lattice rotation in $Zn_3P_2$. The components of the 2D strain tensor and the rigid rotation are plotted with the x-axis in-plane and the y-axis out-of-plane, as shown in (a-c). No significant strain is observed in the whole crystal grain, with strain variations below ±0.1%. The origin of the different strains in the centre of the crystal compared to the top and bottom is unclear. (d) Corresponding lattice rotation map, indicating no perceptible rigid body rotation.

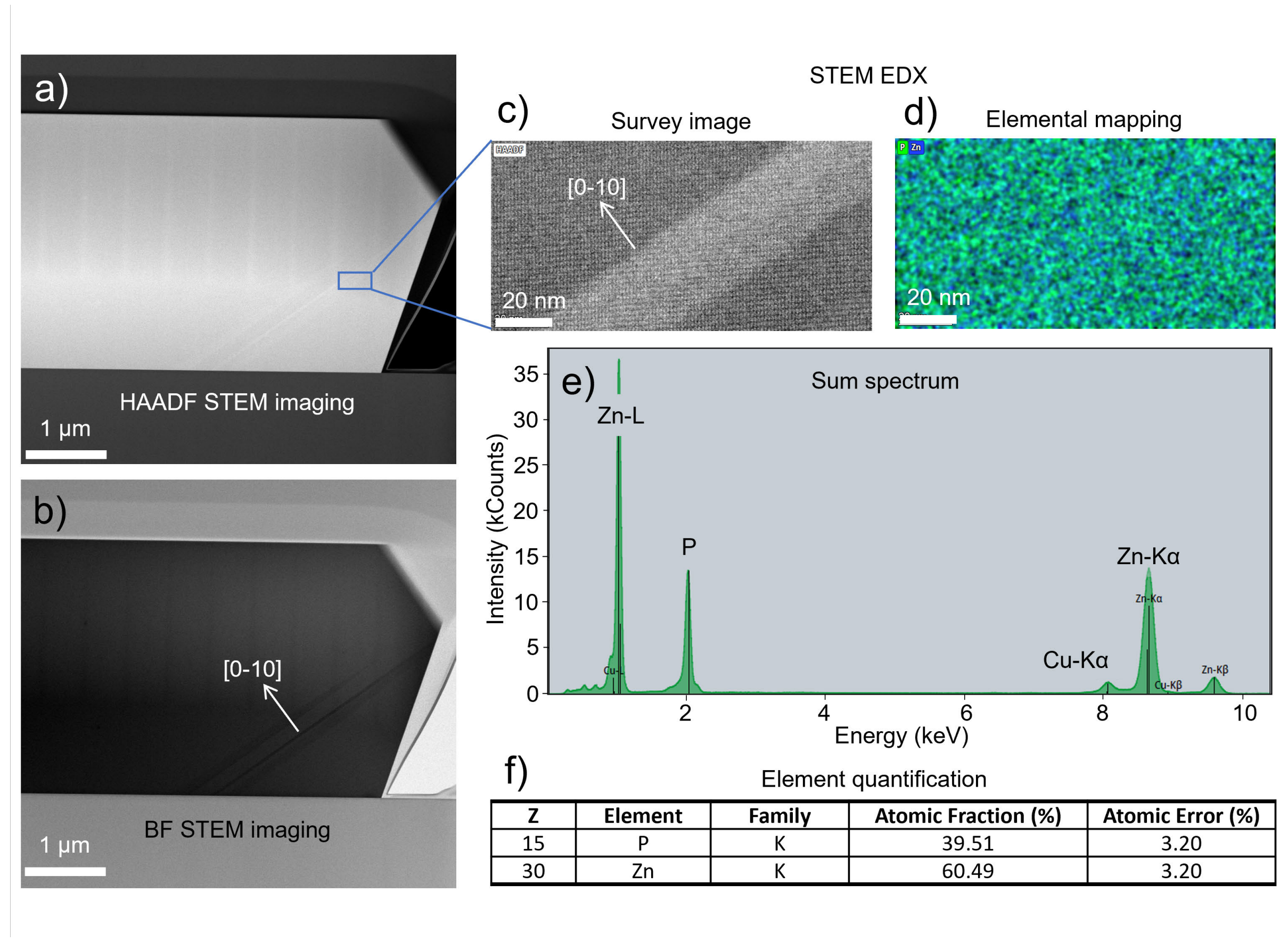

| Z | Element | Family | Atomic Fraction (%) | Atomic Error (%) |
|---|---|---|---|---|
| 15 | P | K | 39.51 | 3.20 |
| 30 | Zn | K | 60.49 | 3.20 |

Figure S8: Summary of the STEM-EDX measurements. (a) shows the HAADF and (b) BF STEM images. (c-f) show the elemental mapping of Zn and P at the position of a (110) APB visible on the survey image. A homogeneous composition is observed. The sum spectrum is quantified, and the stoichiometric composition is measured.

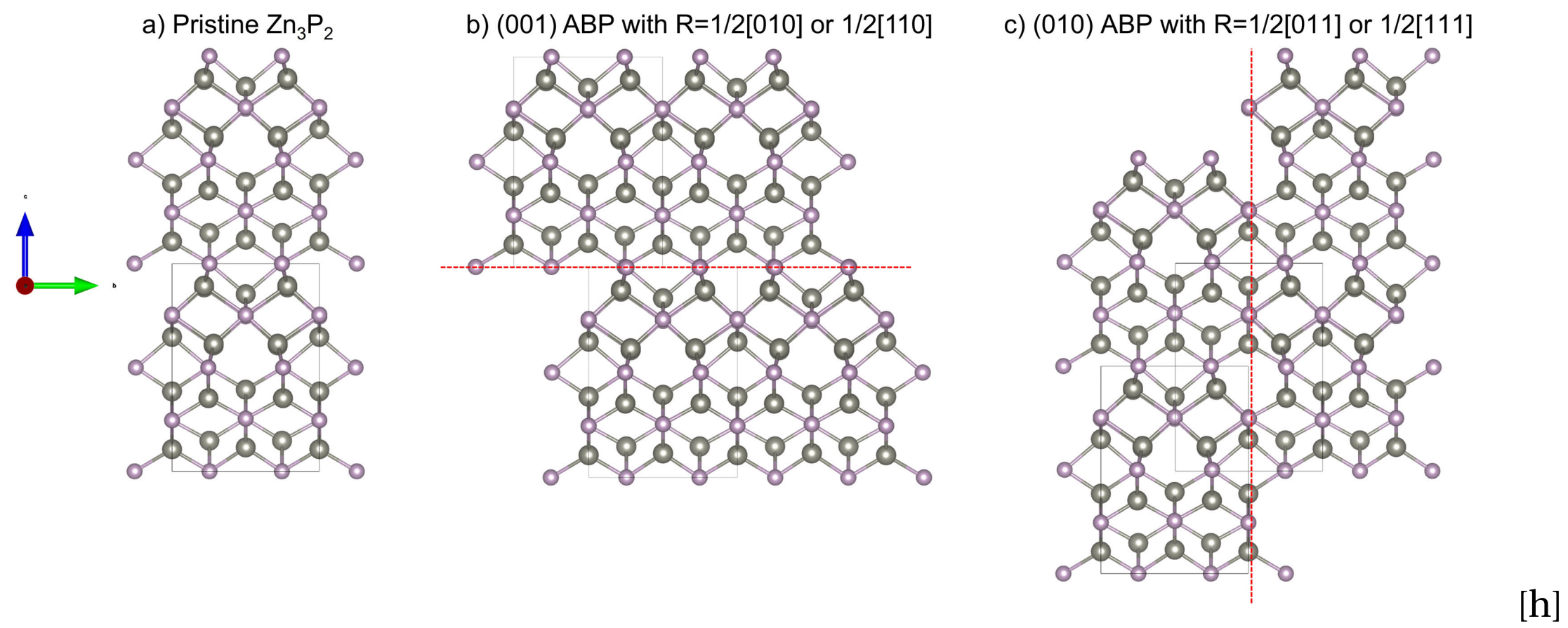

Figure S9: $Zn_3P_2$ p4nmc crystal structure displayed using Vesta software of pristine (a) (001). ABP with displacement vector R=1/2[010] or 1/2[110] (b) and APB with R=1/2[011] or by R=1/2[111] (c).

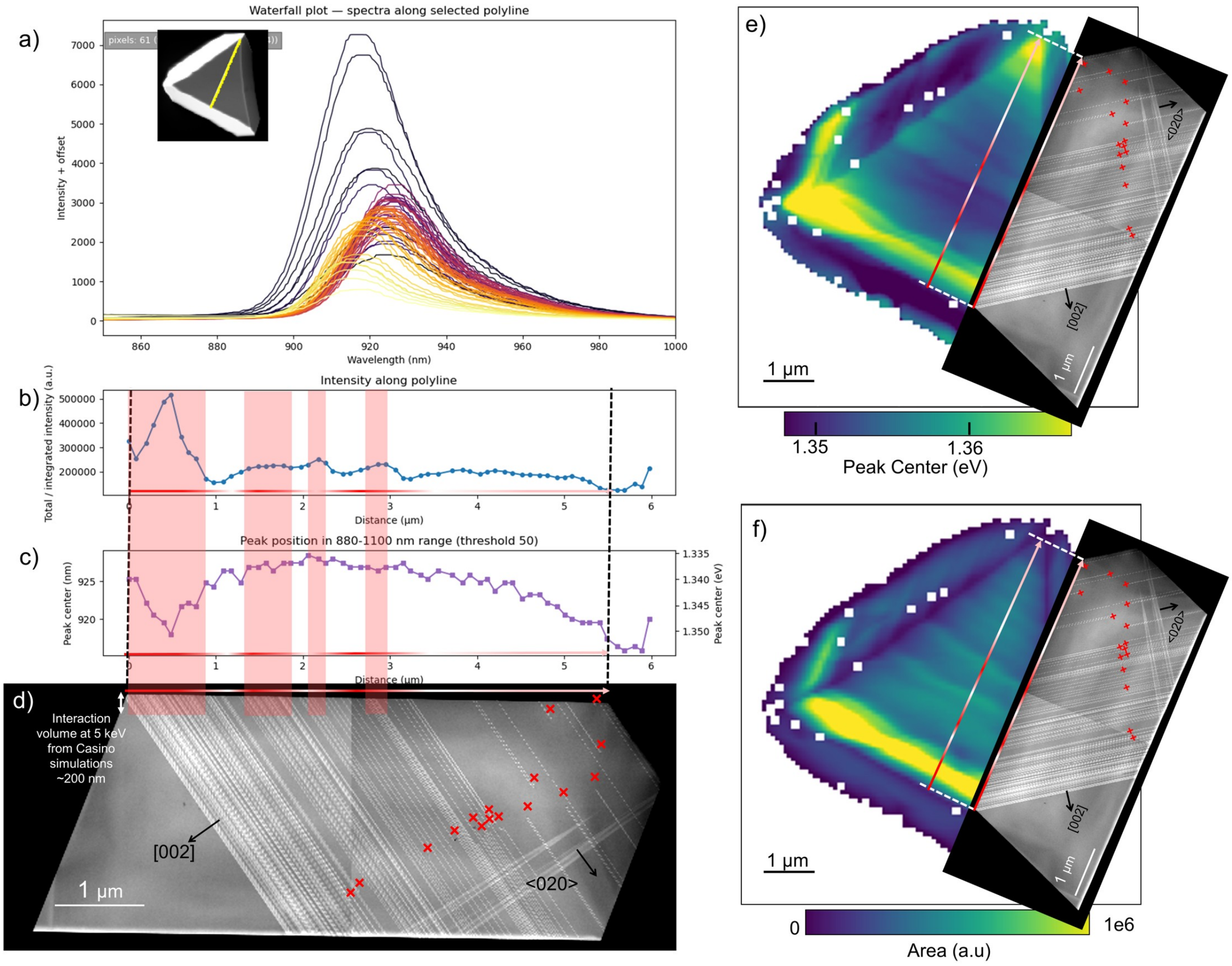

Figure S10: Correlation between cathodoluminescence (CL) emission and antiphase boundary (APB) distribution in a $Zn_3P_2$ grain. CL measurements were acquired at 10 K, 5 keV and 0.5ñA. (a) Waterfall plot of CL spectra acquired along the selected polyline indicated in the inset, positioned at the same location as the TEM lamella extraction. (b) Integrated CL intensity and (c) peak centre position extracted along the same line, showing spatial variations across the grain. Shaded regions highlight areas of increased APB density. (d) Virtual dark-field image extracted from precession 4D-STEM data, indicating the distribution of APBs within the lamella and interaction volume based on Casino simulations.[1] (e) Spatial map of the CL peak centre energy and (f) integrated emission intensity, overlaid with the TEM lamella. A clear spatial correlation is observed between regions of high APB density and variations in both emission intensity and peak energy. While the correspondence is not one-to-one, the CL variations consistently follow the overall distribution of APBs and their local density, indicating a strong, albeit not perfect, relationship between structural defects and optical response.

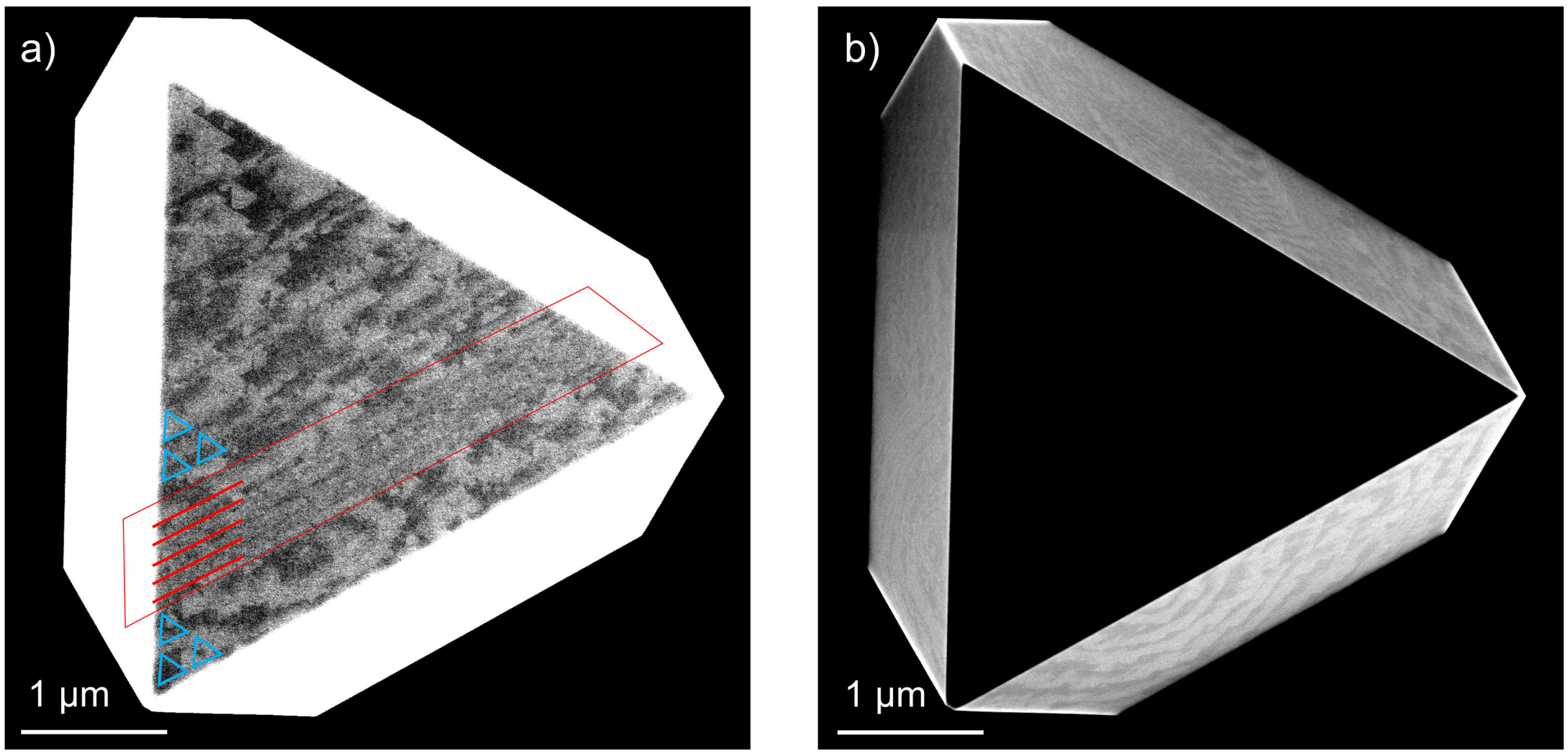

Figure S11: Top-view SEM image of a monocrystalline $Zn_3P_2$ grain. a) and b) show different contrast filtered images to expose structural variation at the surfaces. The red-marked region in (a) exhibits line-like surface texture, different from the rest of the top surface. Although this grain is different from the one analysed by TEM, it is possible that the variation in surface texture is induced by a high density of APBs.

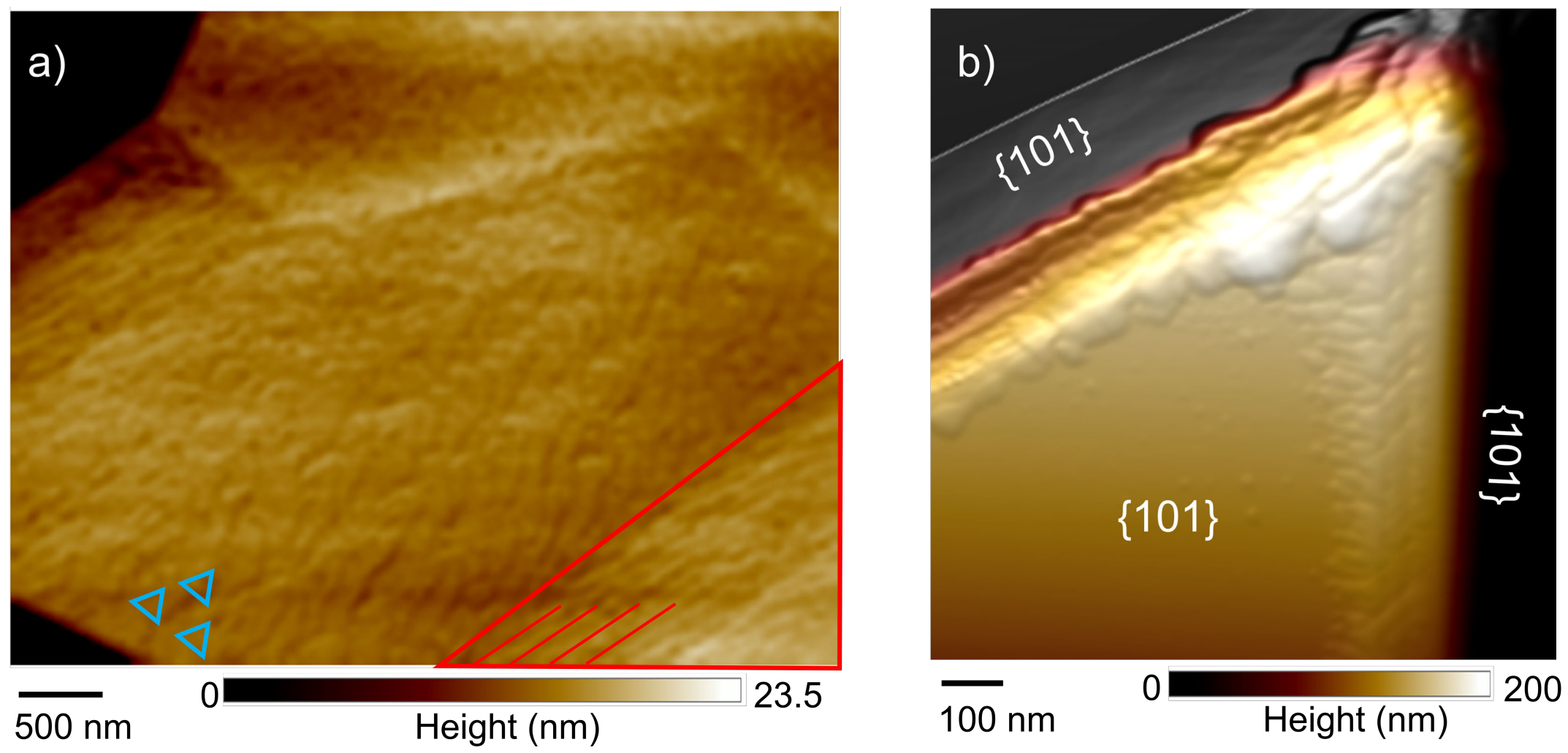

Figure S12: Atomic force micrograph images of $Zn_3P_2$ grown by MOVPE. a) and b) show different regions of grains. The red marked region in (a) exhibits line-like surface texture, similar to the one observed in Fig. S11. b) shows the topography of the edges where the 101 planes merge together.

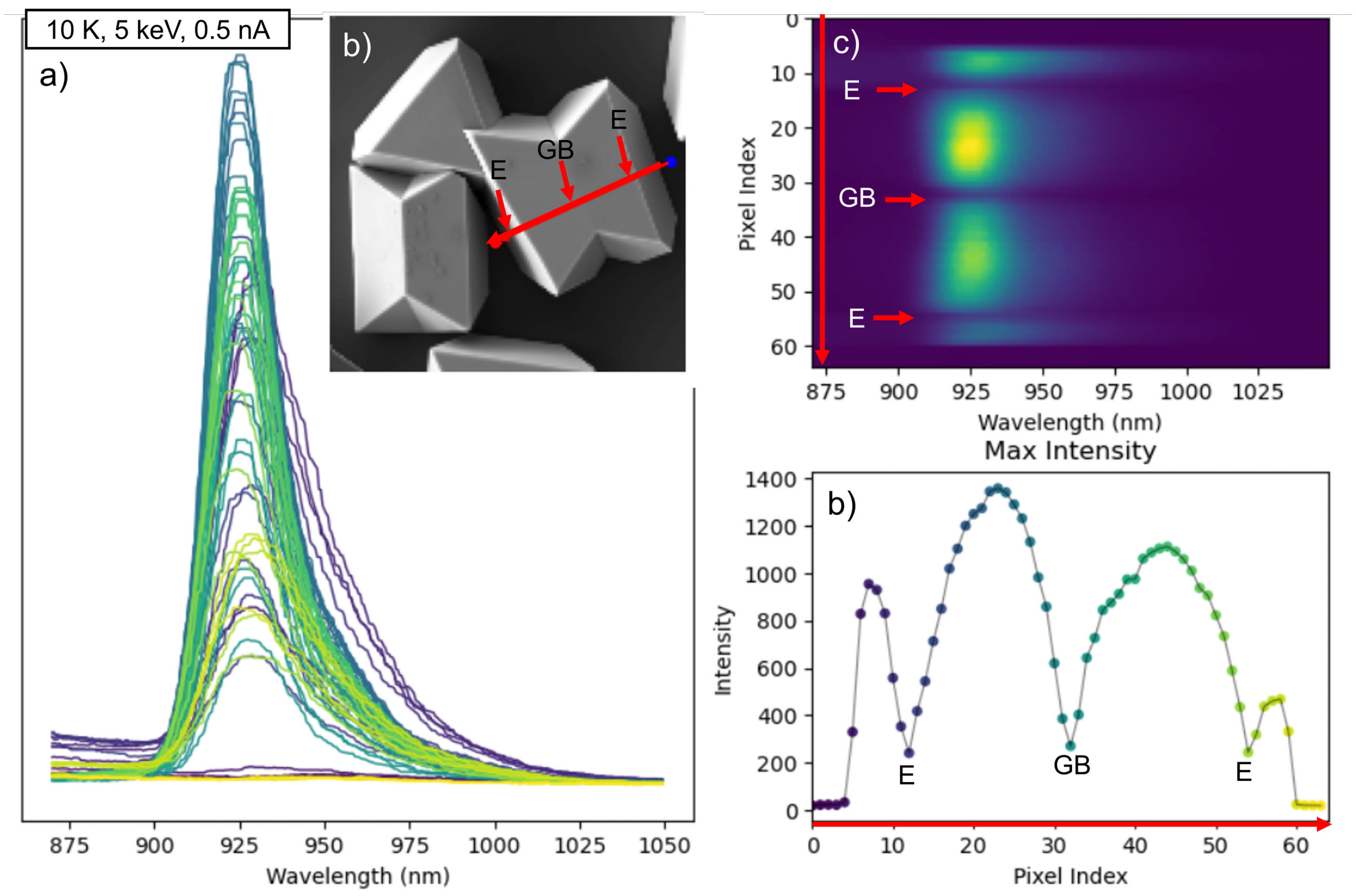

Figure S13: CL measurement of $Zn_3P_2$ grown by MOVPE. a) shows the emission at 10 K along the indicated line in the SEM image in (b). c) shows the spectra of each pixel along the line in a heatmap, with the grain boundary (GB) and edges (E) indicated. d) shows the maximum intensity of the spectrum at each pixel. A clear quenching of the signal is observed at the GB and the edge, which may be related to the rough surface texture at the edge, observed in the AFM images in Fig. S12.

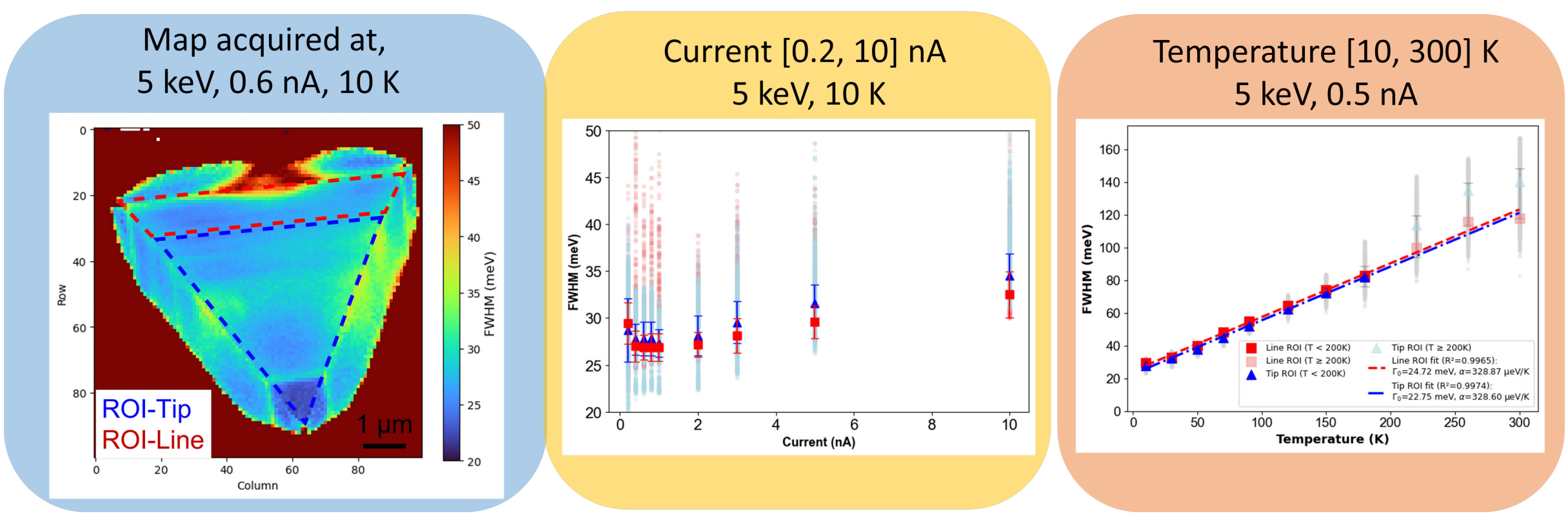

Figure S14: Linewidth corresponding to the fits represented in Fig.**??**. a) Representative FWHM map of the acquisition at 10 K, 5 keV, 0.6 meV. b) FWHM evolution with current and c) with temperature. No significant difference observed between the two ROIs.

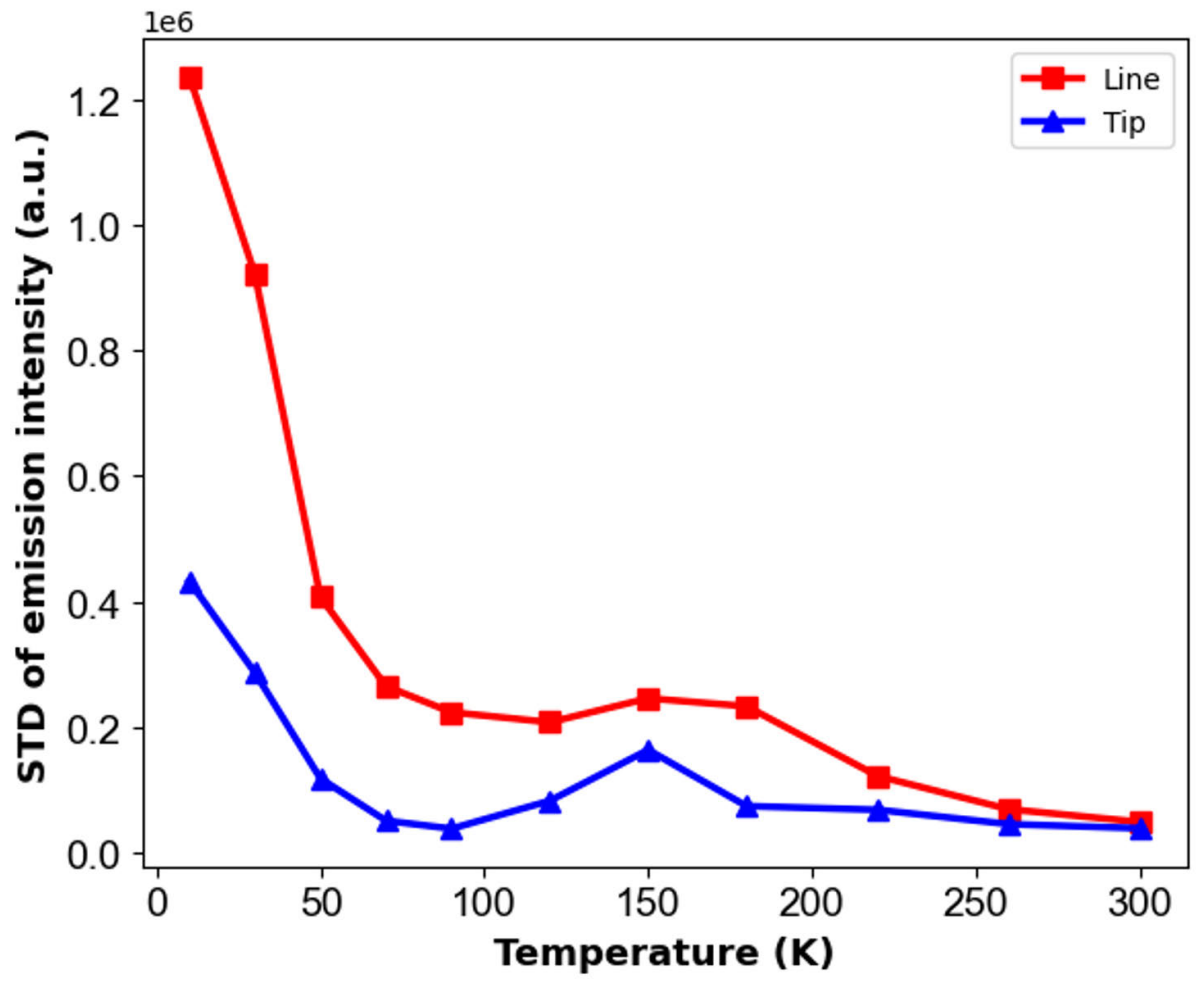

Figure S15: Standard deviation of emission intensity within the two ROIs. A high STD is observed between the different pixel at low T. As the temperature increases, the emission homogenises and the two ROIs become more similar.

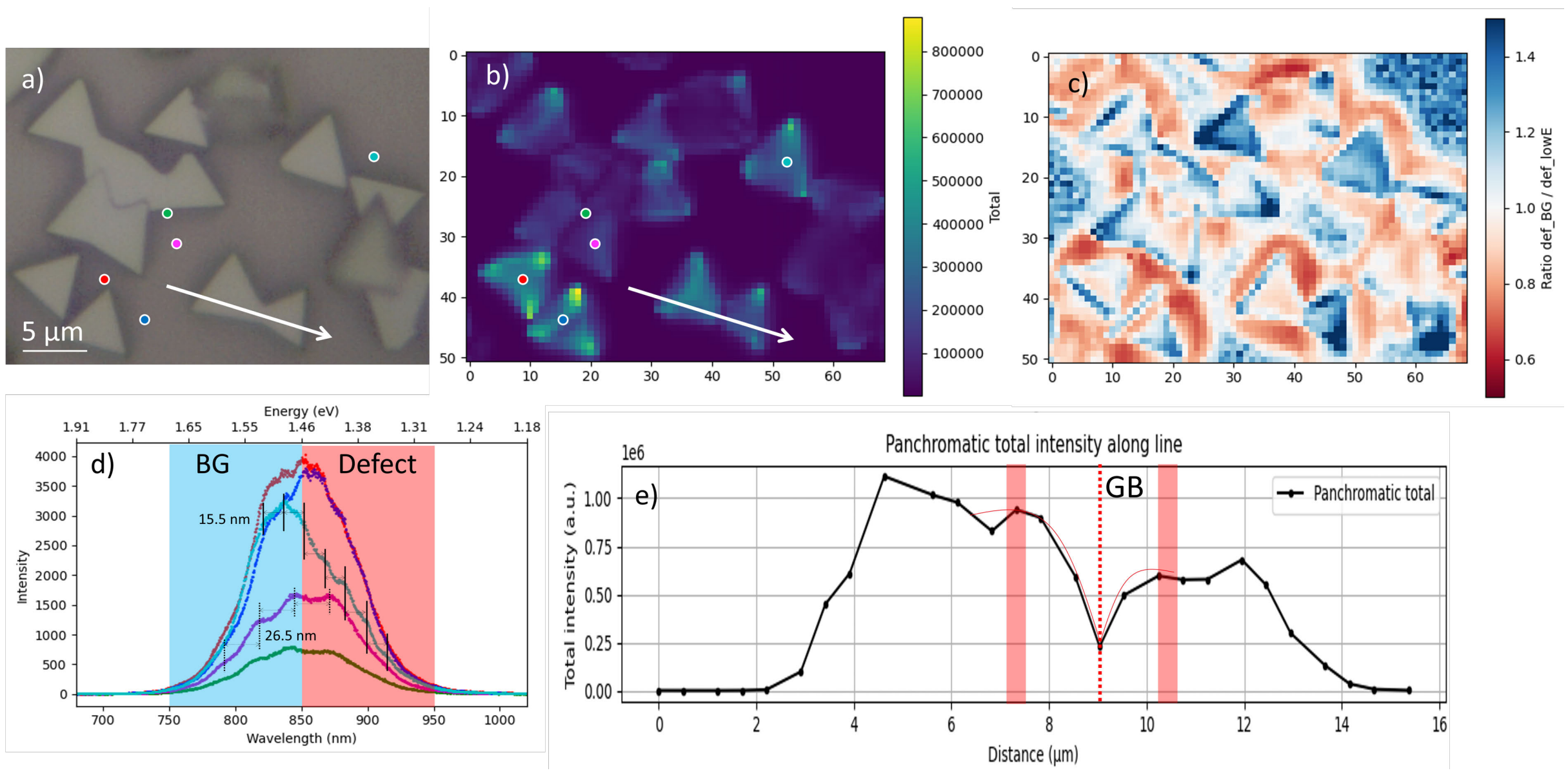

Figure S16: RT-PL map of MOVPE-grown $Zn_3P_2$ acquired with a 532 nm laser, laser power of 180 $\mu$W, 300 l/mm grating and a microscopy objective with NA=0.85. Map contains 50x68 pixel. a) Microscopy image of the mapped zone b) Panchromatic intensity map c) Ratio of the integrated intensity [750-850] nm in blue and [850-950] nm in red. More red appearing regions do have a higher amount of sub-bandgap emission. Large variations in grain colouring are observed, also resembling lines as observed in CL. d) Example spectra at the indicated positions in (a), with the noted integrated ranges in red and blue. Interference patterns are visible in the spectrum, which do, however, not match any dimension of the grain, and therefore remain of unclear origin. e) Panchromatic intensity linescan across a grain boundary, indicated in (b), pointing towards a recovery length in the order of the micrometre.

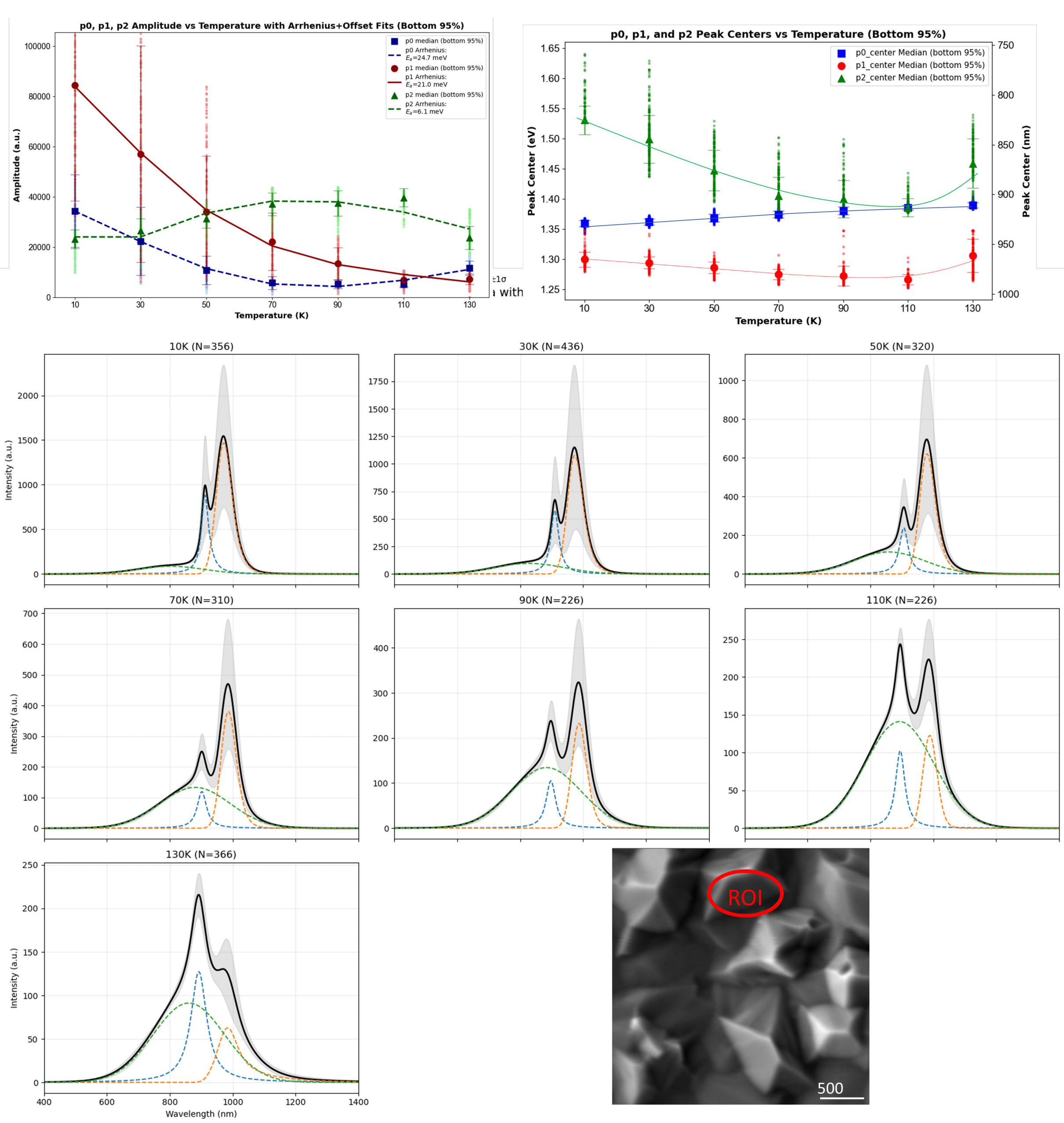

Figure S17: CL emission behaviour of the MBE-grown $Zn_3P_2$ thin film with varying temperature. Arrhenius plot, peak centre evolution and the average spectrum in the ROI at each temperature, with the average fit component p0 (HE),p1 (LE) and p2. A broad Gaussian (green) was fitted above the HE emission (blue), which redshifts with temperature and is broader than the other emissions, likely due to band-to-band or tail-to-tail transitions. The p1 (LE) emission behaves similar with temperature as the reported emission in the MOVPE-grown sample.

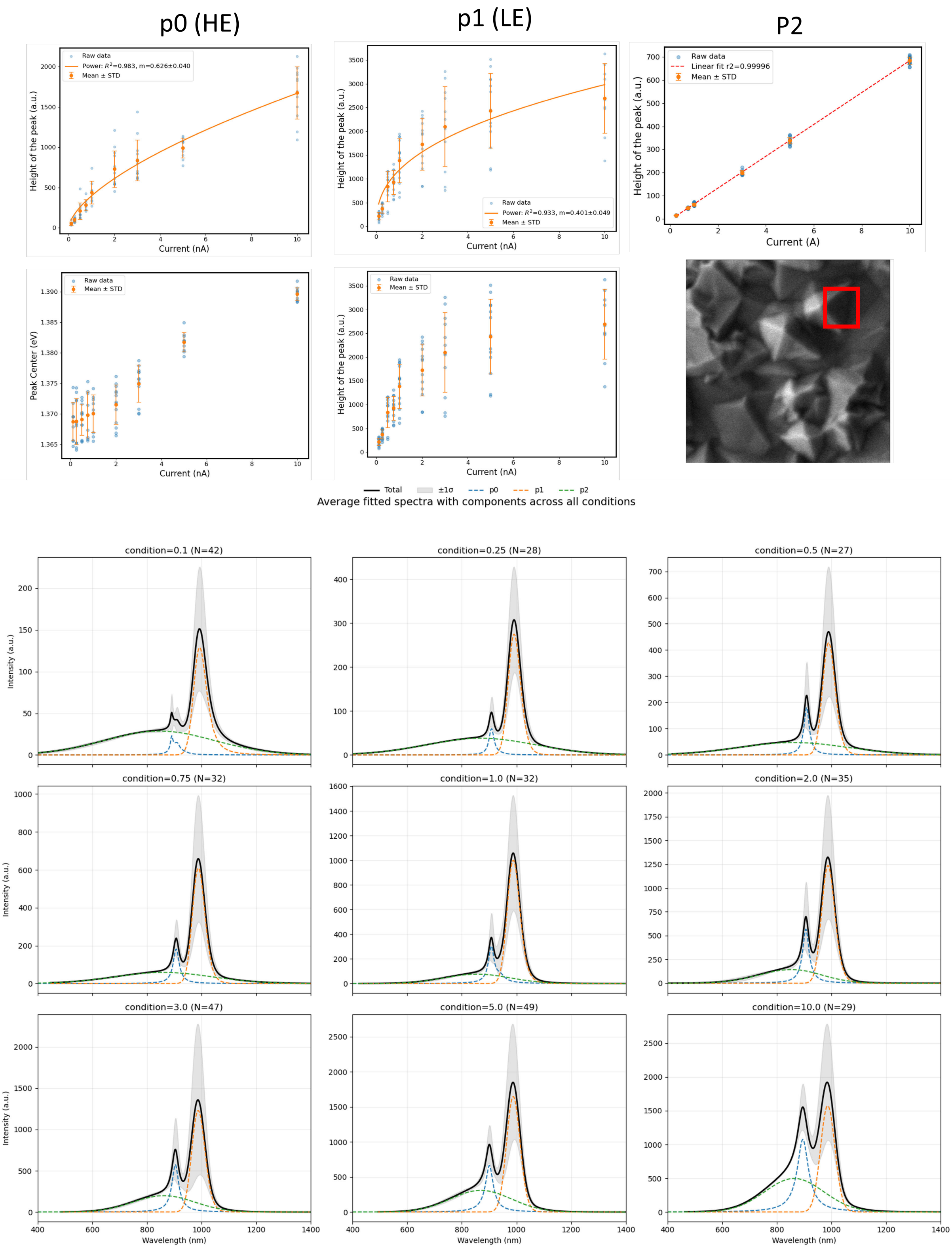

Figure S18: CL emission behaviour of the MBE-grown $Zn_3P_2$ thin film with varying excitation current. For all three peaks, the amplitude evolution is given, indicating saturation for the HE and LE peaks and a linear behaviour for the p2 peak, further supporting bandgap-related emission of p2. p1(LE) behaves similar with current as the reported emission in the MOVPE-grown sample.

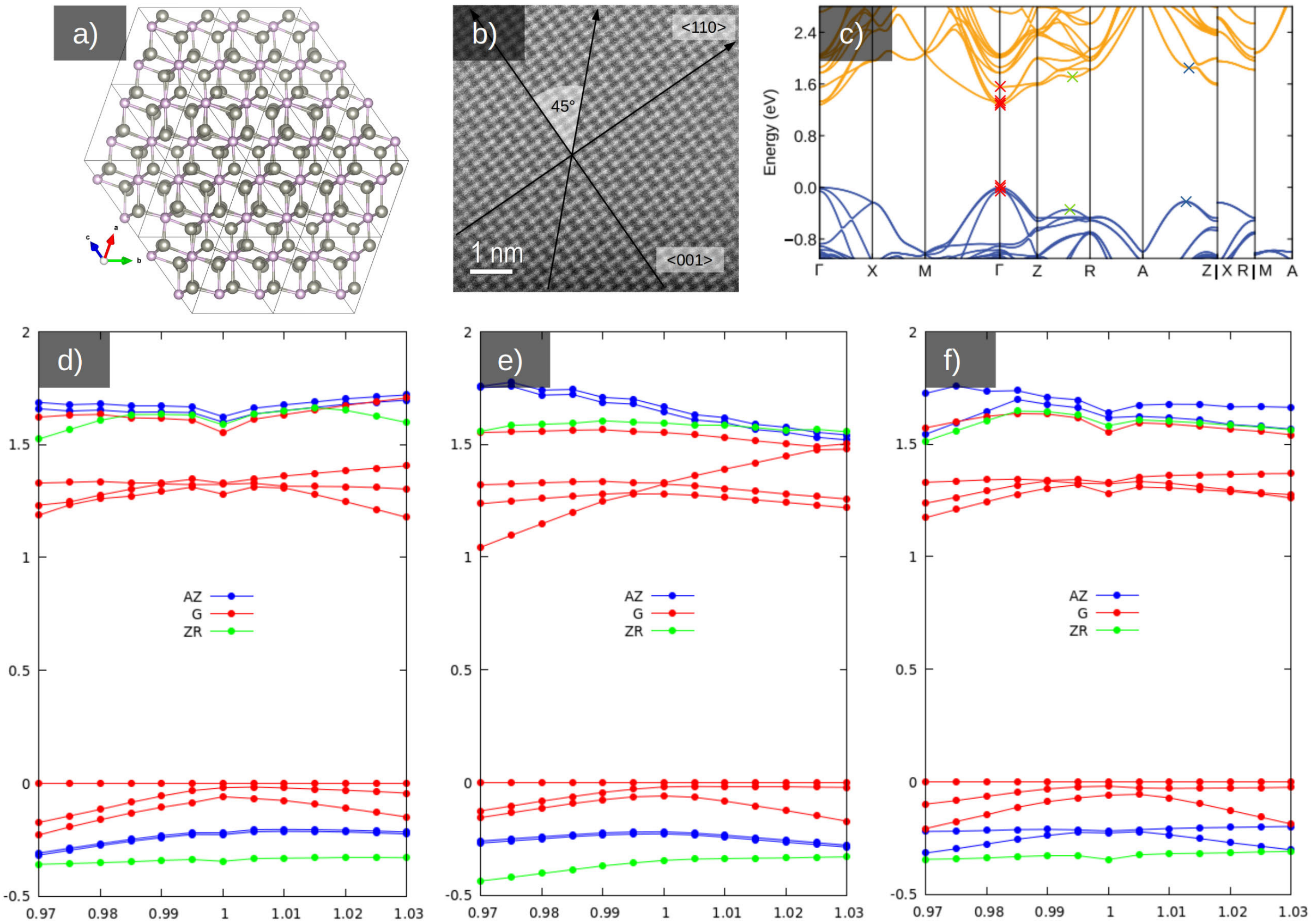

Figure S19: Density functional theory (DFT) analysis of strain effects in $Zn_3P_2$. (a) Model crystal structure projected along [$\bar{1}$11] with the <101> vector pointing upwards. (b) High-resolution TEM image in matching projection and orientation. The three indexed directions illustrate the three directions in which uniaxial strain was applied. (c) Calculated electronic band structure along high-symmetry directions. Crosses mark electronic states whose evolution under strain was investigated. (d–f) Evolution of selected electronic states under uniaxial strain applied along (d) <001>, (e) 45° between <001> and <110>, and (f) <110> directions. The energy shifts of valence and conduction band states reveal a pronounced anisotropy in the strain response, indicating direction-dependent modifications of the band edges and suggesting that the electronic structure is sensitive to local lattice distortions.

# References

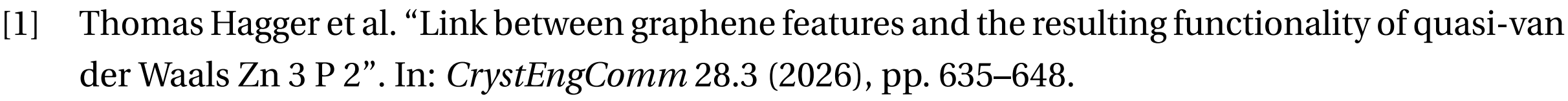